# Pore-level Quantitative Structure-Activity Relationship (QSAR) for Water Permeation Rate in Aquaporins

Juan José Galano-Frutos[1,a,†,*], Luca Bergamasco[2,a,*], Paolo Vigo[2], Matteo Morciano[2], Matteo Fasano[2], Davide Pirolli[1], Eliodoro Chiavazzo[2] and Maria Cristina de Rosa[1]

[1] Istituto di Scienze e Tecnologie Chimiche "Giulio Natta" (SCITEC) - National Research Council (CNR), Via Largo Francesco Vito 1, 00168 Rome, Italy
[2] Department of Energy, Politecnico di Torino, C.so Duca degli Abruzzi 24, 10129 Turin, Italy

[†] Present address: Certest Biotec S.L., 50840 San Mateo de Gallego, Zaragoza, Spain

[a] These authors contributed equally to this work.
*** Correspondence:** J.J. G.-F. jjgalano@certest.es; L.B. luca.bergamasco@polito.it

**ABSTRACT**

Aquaporins (AQPs) and aquaglyceroporins (AQGPs) play a crucial role in regulating water transport and solute selectivity across biological membranes. Besides their biological relevance, AQPs have attracted growing interest as models for the design of next-generation biomimetic membranes for water filtration. In this work, we present a pore-level Quantitative Structure–Activity Relationship (QSAR) approach that relates structural and physicochemical pore descriptors with experimentally reported water permeation rates across a set of AQ(G)Ps with high-resolution 3D structures. This data-driven methodology, presented here as a proof of concept, introduces a multi-feature framework for determining pore descriptors associated with water transport efficiency in AQ(G)Ps. Applied to two compiled permeation rate datasets, this framework recapitulates determinants previously reported in single-feature studies, while also highlighting additional pore descriptors that emerge as relevant in a multi-variable context. The insights gained through this approach may, in perspective, contribute to advancing the rational design of AQP-based filtration devices and to deepening the molecular understanding of the function of these valuable macromolecules in health and disease.

## 1. INTRODUCTION

The maintenance of ionic homeostasis and water levels within the cell and within different organisms' tissues and compartments is a function that has been linked to a protein family called aquaporins (AQPs), which includes aquaglyceroporins (AQGPs). Initially referred to as Channel-forming Integral Protein (CHIP) [1–3] or Major Intrinsic Proteins (MIPs), AQ(G)Ps are 28-35 kDa integral membrane protein channels, which enable the transport of water across cells with a high efficiency and a unique selectivity. Typically assembled as tetramers (Fig. 1a), each subunit forms an integral membrane pore, characterized by six transmembrane spanning domains, with cytosolic amino and carboxy termini (Fig. 1b and 1c). Only pure water and small uncharged solutes, e.g. glycerol, urea, metalloids, carbon dioxide, oxygen, nitrogen, hydrogen peroxide and ammonia, are allowed to pass through the pore [4–7]. This selectivity is due to an electrostatic barrier located at the center of the channel, which prevents the passage of protons and other charged molecules [8] (Fig. 1b). Phenomena like the survival of some plants in extreme environments, characterized by saline-alkaline soils, high or low temperatures, heavy metal toxicity and drought are owed to adaptations developed by these organisms thanks to the presence of AQ(G)Ps [9–12]. AQ(G)Ps have also been associated with cell volume regulation, transepithelial fluid transport, cell migration and proliferation, neuroexcitation, among other physiological events [13–15], and thus linked to various non-infectious and infectious diseases in humans [3,13,14,16,17], highlighting the importance of their study across a variety of research and development fields.

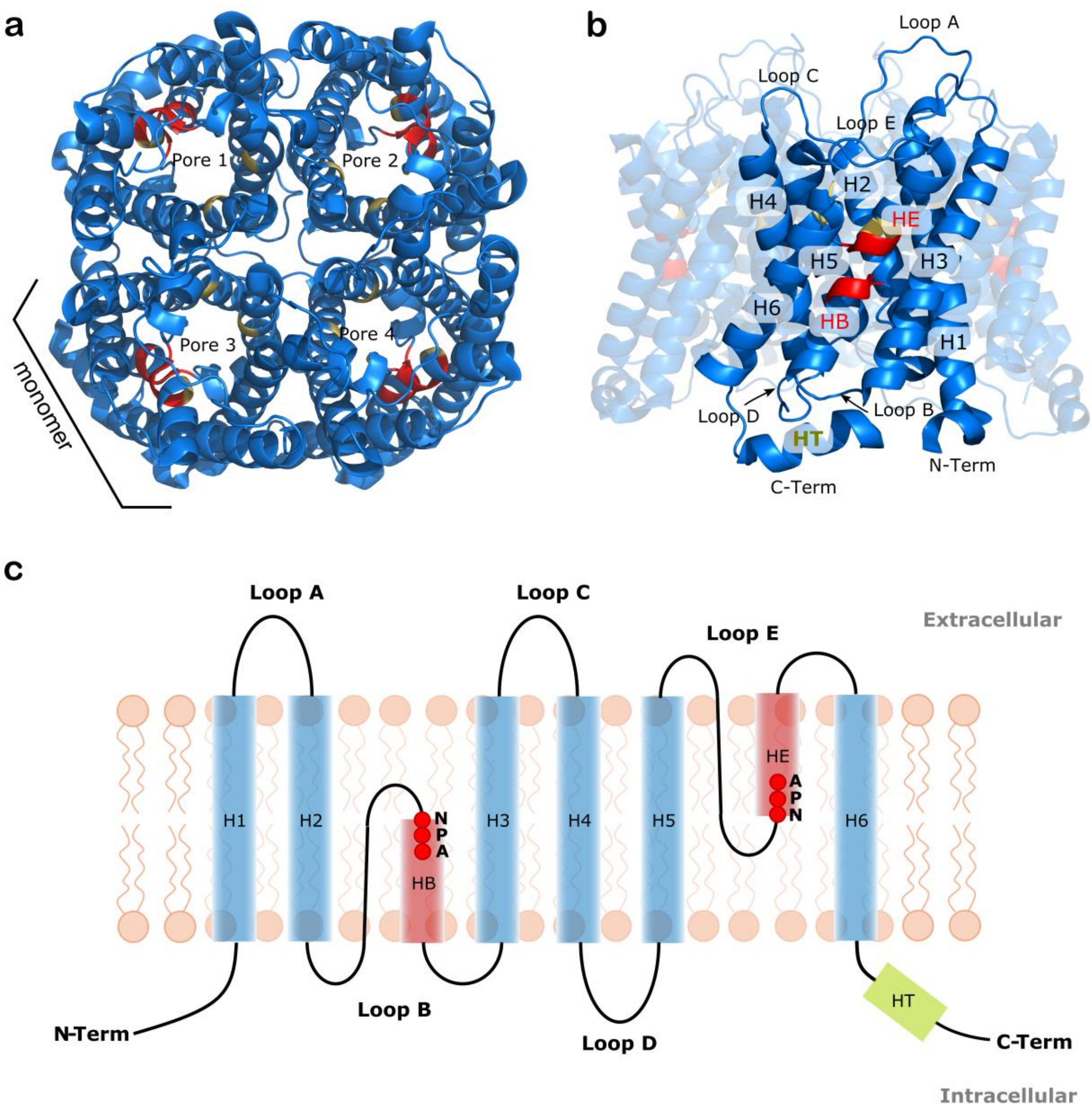

**Figure 1.** Structural features of mammalian AQ(G)Ps. (a) Tetrameric structure of hAQP1 (PDB ID 7UZE [18]), viewed from the extracellular side, with the pores indicated on each monomer. (b) Side view of hAQP1 highlighting the main features of the pore for one monomer. The two NPA motifs are highlighted in red, and the selectivity filter (SF) motif is highlighted in yellow. (c) Schematic representation of the mammalian AQ(G)P channel topology. Each monomer comprises six transmembrane helices (H1–H6) connected by five loops (A–E) and an additional C-terminal short amphipathic helix (HT). Loops B and E fold back into the membrane, forming two half-membrane spanning helices (HB and HE) that harbor the two asparagine–proline–alanine (NPA) motifs (red colored circles). The N- and C-termini are located in the cytoplasm, and the amphipathic helix (HT) is a common protein–protein interaction site.

There are more than 350 known aquaporins, with 13 subfamilies found in mammalian, namely AQP0-12. Mammalian AQ(G)Ps are usually divided into three sub-classes, one containing proteins that allow only water molecules flow, namely AQP0, 1, 2, 4, 5, 6 and 8, which are referred to as orthodox aquaporins, another one that allows the transport of water and other small solutes, like glycerol or urea, that includes AQP3, 7, 9 and 10, referred to as aquaglyceroporins (AQGPs), and a more recently discovered class that has comparatively lower sequence homology (about 20%) with the orthodox aquaporins, that includes AQP11 and AQP12, referred to as unorthodox or superaquaporins [17,19,20]. A higher AQPs diversity is present in plants, where there are five homologous AQPs subfamilies, namely plasma membrane intrinsic proteins (PIP), tonoplast intrinsic protein (TIP), small basic intrinsic protein (SIP), nodulin-26 like intrinsic protein (NIP) and X intrinsic proteins (XIP) [10]. Amphibians [21], bacteria [22] and fungi [23] have also been reported to carry AQPs.

Thanks to their inherent functional characteristics, aquaporins are used for developing next-generation ultra-permeable membranes [24,25]. Aquaporin-based biomimetic membranes (ABMs) [26] have demonstrated remarkable promise both in water desalination [27,28] (see reviews in Ref. [29–33]) and wastewater treatment [34–36]. Their application for these and other related purposes has proven to be a valuable solution [37–39] as an alternative to conventional reverse osmosis (RO) [40,41] or thermal desalination techniques [42,43], offering potential energy and cost savings [39]. Hélix-Nielsen [31] and Beratto-Ramos et al. [36] emphasized the necessity of adopting novel optimization and design strategies to enhance ABM technology further. In close alignment with this imperative, the past 15 years have witnessed a significant surge in the development of bio-inspired artificial water channels (AWCs), which has been driven by the ambition to fabricate synthetic counterparts of AQ(G)Ps (see Ref. [44–46], and recent applications in Ref. [47,48]), marking a notable frontier in scientific exploration.

In connection with these challenges, significant efforts—spanning both experimental and computational domains—have been dedicated to unraveling the working principles governing the permeation and selective capabilities of AQ(G)Ps [19,49–60]. Most of them, however, have predominantly focused on individual features of these proteins [61], as for instance, the amino acid compositions of the NPA and SF motifs, the pore gating [59], the pore shape and size [60,61], the putative number of hydrogen bonds formed along the pore lining [57], or the number of positive charges at the pore mouth [58], to name a few. For instance, it has been shown that AQ(G)Ps' water permeability and high solute rejection occur through a narrow hydrophilic constriction, typically around ~3 Angstroms, which facilitates the formation of a water single-file wire [62]. This process is characterized by hydrogen bonding among water molecules [50,63], alongside bonding and polarizing interactions with the pore lining [57,58,62]. Also, amino acids positions determinative of AQP vs. AQGPs selectivity, and those involved in auto-regulatory events linked to post-translational modifications have been unveiled through a variety of experimental methods, including site-directed mutagenesis, chimeric domain swaps, membrane permeability assays, electron crystallography, X-ray crystallography, and molecular dynamics simulations [19,64,65]. Despite such progress, a precise and comprehensive understanding of the permeation properties, selectivity and mechanisms exhibited by natural AQ(G)Ps remains incomplete [66].

Quantitative Structure-Activity Relationship (QSAR) (or Quantitative Structure-Property Relationship, QSPR) modeling, is one of the most extended in silico tools employed in medicinal chemistry [67]. In QSAR, features such as solubility, acidity, and polarity are computed as molecular descriptors and then applied to determine which of them can be used to predict measurable tasks, like toxicity, drug-likeness, or enzyme binding. The application of QSAR to model the activities or properties of peptides and proteins is an emerging field that is not yet as well established as it is for small molecules, due to the added complexity of these systems. Nevertheless, this approach has proven useful for uncovering hidden relationships that can be exploited to guide the design and optimization of the properties of systems of interest [68]. For valuable insights into the current state, development, and optimization of molecular descriptors to be used in QSAR studies on peptides and proteins, readers may refer to the following works: Ref. [69] (a review paper) and Ref. [70] (a research article).

Although well-established, QSAR methods appear to be still unexplored in the context of AQ(G)Ps. Thus, here we present—as a methodological proof of concept—a pore-level QSAR approach that combines available structural and permeability data to reveal key, and potentially novel, determinants of the filtering capabilities of these remarkable biomolecules. The proposed methodology could provide deeper insights at the molecular level, offering a better understanding of the AQ(G)Ps' functioning, and creating opportunities for the rational improvement of their properties and for targeted interventions in disease-related contexts.

## 2. MATERIALS AND METHODS

### 2.1. AQPs with available 3D structure

A search for proteins of the MIP family was performed on two different databases, namely PROSITE [71,72] and InterPro [73,74]. Proteins with experimental 3D structures available were subsequently retrieved from the RCSB PDB archive [75].

The PROSITE database [71,72], returned 285 entries (with accession code PS00221 related to MIP family). Out of the true positive sequences (264), 53 structure entries were available on the PDB archive according to this database. The InterPro database [73,74], which contained all the structure entries reported in PROSITE, returned (with accession code IPR000425 related to MIP family) 73 different PDB structures. This number of structures represented 22 individual aquaporins. In cases where multiple structures were available for a protein, the one with the highest structural resolution among those with canonical sequences (no mutations) was selected as its representative structure for the present study. An overview of these 22 AQ(G)Ps and their respective PDB IDs is reported in Table 1.

**Table 1.** Overview of AQPs with available 3D structure and selected PDB entries.

| AQ(G)P (short name) | PDB ID | Source organism | Ref.[a] |
|---|---|---|---|
| Aquaporin 0 (bAQP0) | 1YMG | *Bos taurus* | [76] |
| Aquaporin 0 (oAQP0) | 2B6O | *Ovis aries* | [77] |
| Aquaporin 1 (bAQP1) | 1J4N | *Bos Taurus* | [78] |
| Aquaporin 1 (hAQP1) | 7UZE | *Homo sapiens* | [18] |
| Aquaporin 1 (aAqp1) | 7W7S | *Anabas testudineus* | [79] |
| Aquaporin 1 (kAqp1) | 3ZOJ | *Komagataella pastoris* | [80] |
| Aquaporin 2 (hAQP2) | 4NEF | *Homo sapiens* | [81] |
| Aquaporin 4 (hAQP4) | 3GD8 | *Homo sapiens* | [82] |
| Aquaporin 4 (rAQP4) | 2D57 | *Rattus norvegicus* | [83] |
| Aquaporin 5 (hAQP5) | 3D9S | *Homo sapiens* | [84] |
| Aquaporin 7 (hAQP7) | 6QZI | *Homo sapiens* | [85] |
| Aquaporin 10 (hAQP10) | 6F7H | *Homo sapiens* | [86] |
| Aquaporin (SoPIP2;1) | 1Z98 | *Spinacia oleracea* | [87] |
| Aquaporin M (mAqpM) | 2F2B | *Methanothermobacter marburgensis* | [88] |
| Probable aquaporin M (aAqpM) | 3NE2 | *Archaeoglobus fulgidus* | [89] |
| Aquaporin NIP2;1 (OsNIP2;1) | 7CJS | *Oryza sativa japonica* | [90] |
| Aquaporin TIP2;1 (AtTIP2;1) | 5I32 | *Arabidopsis thaliana* | [91] |
| Probable aquaporin PIP2;4 (AtPIP2;4) | 6QIM | *Arabidopsis thaliana* | [6] |
| Aquaporin Z (eAqpZ) | 1RC2 | *Escherichia coli* | [92] |
| Aquaporin Z 2 (aAqpZ2) | 3LLQ | *Agrobacterium fabrum* | [93] |
| Aquaglyceroporin (pAqgp) | 3C02 | *Plasmodium falciparum* | [94] |
| Glycerol uptake facilitator protein (eGlpF) | 1FX8 | *Escherichia coli* | [62] |

[a] Reference articles for structures 3LLQ and 3NE2 are not yet available. In these cases the references provided are the links to the entries in the RCSB Protein Data Bank [75].

### 2.2. Training Set and AQ(G)Ps' Permeation Rate

We reviewed the literature for osmotic unitary (single-channel) water permeation rates ($p_f$)—encompassing both experimental and molecular dynamics (MD)-based approaches—reported for the 22 AQ(G)Ps with available experimental structure (see Table S1 and Fig. 2). Permeation data for other AQ(G)Ps lacking resolved 3D structures, and therefore not included in Table 1, are also provided in Tables S1 and S2 and in Fig. 2 (see details below).

Experimental data was not found for some AQ(G)Ps listed in Table 1, for which only *in silico* data is reported. For some AQ(G)Ps, we could not find any data, or the reported permeation rates were not provided as single-channel water permeabilities (i.e., $p_f$) but rather as global osmotic permeability coefficient ($P_f$) or other metrics. In cases such as hAQP7 [95–97] and hAQP10 [86], we were unable to convert $P_f$ into $p_f$ due to the lack of required data for such a transformation. In the case of hAQP5, the functional analysis carried out by Horsefield *et al.* [84] relied on a comparative light-scattering–based approach, for which only rate constants ($k$, in $s^{-1}$) were reported; consequently, we were not able to retrieve the $p_f$ values.

Importantly, Wachlmayr *et al.* [56] recently revisited the stopped-flow methodology underlying many historical measurements of osmotic water permeability ($P_f$) and unitary water permeability ($p_f$). They argued that a significant subset of older reports is systematically biased because $P_f$ was often obtained by fitting light-scattering traces to a single exponential and inserting the resulting time constant into approximate formulas that are only valid under restrictive conditions. To address this, they compare the main analysis strategies and promote an empirically corrected analytical-solution-based exponential approach (E-ASF) that retains the practical simplicity of exponential fitting while reproducing permeabilities consistent with the proper analytical solution. Crucially for legacy datasets, they provide osmolarity-based correction factors that can be applied even when the original traces are unavailable, and show that many published $p_f$ values should be revised downward (typically to ~0.25−0.5 of the originally reported value), in agreement with independent numerical re-fits [56]. Here, we compiled—from Ref. [56]—the corrected unitary water permeabilities (hereafter $p_f^{corr}$) for 8 of AQ(G)Ps listed in Table 1 (i.e., those with experimentally resolved 3D structures), as well as for 4 additional AQPs lacking 3D structures, which are used as an external validation set in the pore-level QSAR approach presented here (see Table S2 and Fig. 2).

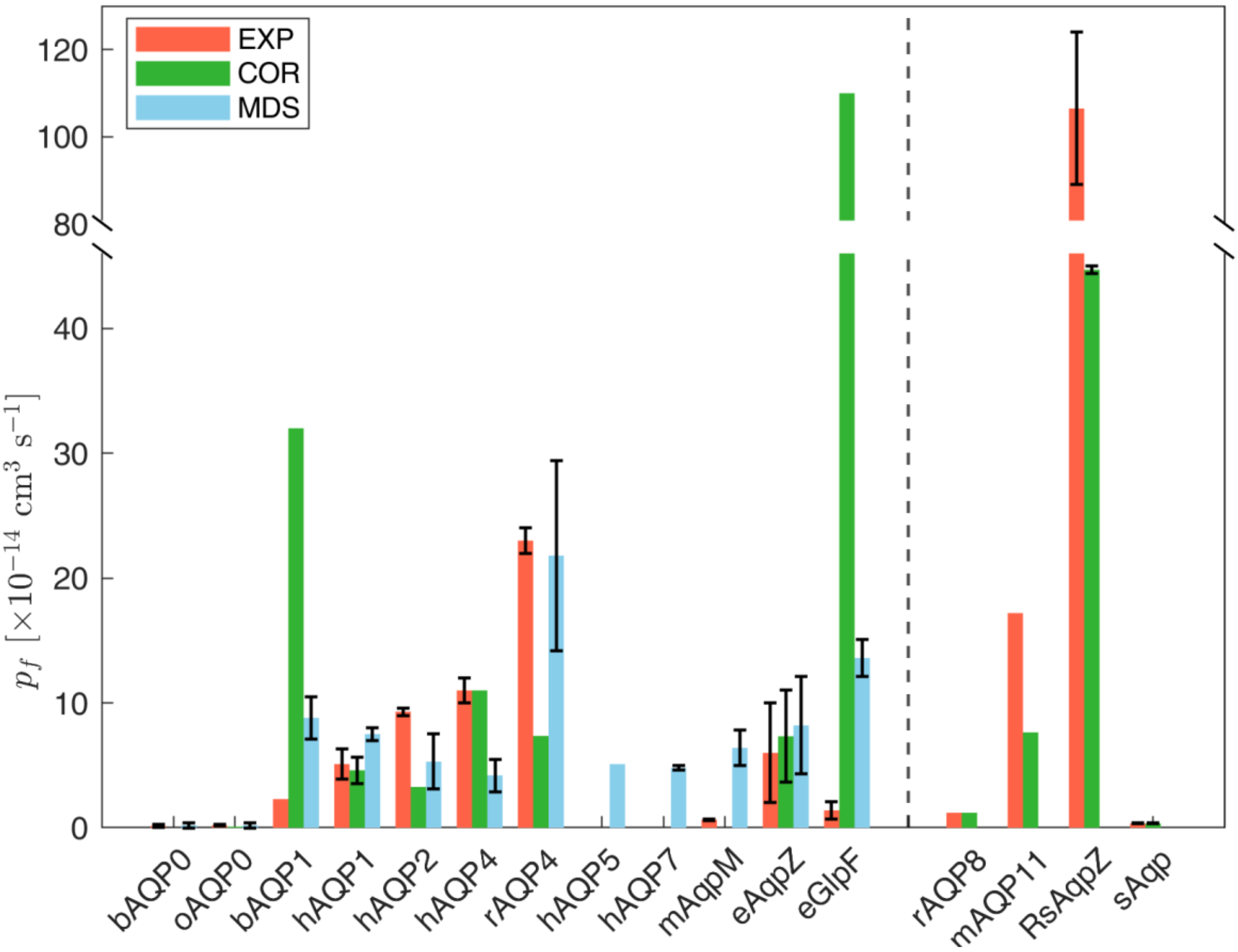

**Figure 2. Osmotic unitary water permeation rates ($\times 10^{-14}$ cm$^3$·s$^{-1}$).** Red bars correspond to individual $p_f$ measurements or averages, as compiled in Table S1. Green bars correspond to corrected unitary water permeabilities ($p_f^{corr}$ [56]), as compiled in Table S2. Cyan bars correspond to $p_f$ values (individual or averages) obtained from molecular dynamics simulations (MDS; as compiled in Table S1). AQPs and AQGPs to the left of the vertical dashed line correspond to those with experimentally resolved 3D structures (listed in Table 1). Within this subset, experimental $p_f$ values could not be found or derived for hAQP5 and hAQP7, and the compiled $p_f^{corr}$ values correspond to corrections or averages derived strictly from low-temperature (≤15°C) measurements (Table S2). AQPs to the right of the dashed line lack resolved 3D structures and were used as an external validation set in the pore-level QSAR approach presented here. The experimental methods employed, along with the main conditions and simulation setups used to derive the compiled values, are summarized in Table S1. When more than one value is available—regardless of the method used—the bar represents the mean value and the black vertical lines indicate the standard error of the mean. Otherwise, the black vertical line denotes the experimental or computational uncertainty—when reported—associated with the individual measurement or calculation.

Since equilibrium MDS-based approaches, mostly used in earlier modeling of AQ(G)Ps, have well-recognized limitations [98], and because MDS-derived osmotic unitary water permeabilities do not appear to align well with either non-corrected or corrected experimental estimates (see Fig. 2), we did not consider MDS-derived permeation rates for the pore-level QSAR framework presented here. We therefore carried out the proposed modelling separately for two subsets of structurally resolved AQ(G)Ps listed in Table 1: (i) those for which experimental $p_f$ values are available (individual measurements or averages thereof, as compiled in Table S1), comprising 10 AQ(G)Ps: bAQP0, oAQP0, bAQP1, hAQP1, hAQP2, hAQP4, rAQP4, mAqpM, eAqpZ and eGlpF (see Table S1 and Fig. 2); and (ii) those for which corrected unitary permeabilities, $p_f^{corr}$, are reported by Wachlmayr *et al.* [56], comprising 8 AQ(G)Ps: oAQP0, bAQP1, hAQP1, hAQP2, hAQP4, rAQP4, eAqpZ and eGlpF (see Table S2 and Fig. 2). This second set of AQ(G)Ps with available $p_f^{corr}$ values is indeed a subset of the first one.

When compiling the $p_f^{corr}$ data from Ref. [56], we followed the authors' recommendation that osmotic permeabilities should preferably be measured at low temperature (<10 °C) to minimize systematic errors. Thus, we applied a temperature-based filtering to maximize internal consistency of the corrected permeability dataset: measurements performed above 10 °C, as well as entries with unreported temperature were discarded (with the sole exception of sheep AQP0, for which only one value at 15 °C was available and was retained). In this context, in cases where multiple low-temperature permeabilities are reported for the same AQ(G)P, we averaged their $p_f^{corr}$ values to obtain a single representative corrected permeability per aquaporin (Table S2).

For QSAR modeling, unitary permeabilities were expressed on a common numerical scale (×$10^{-14}$ $cm^3 \cdot s^{-1}$, after which the $10^{-14}$ factor was removed prior to log-transformation. We thus used $ln(p_f)$ or $ln(p_f^{corr})$ as the response variable in the QSAR modeling. This log transformation was necessary because $p_f$ and $p_f^{corr}$ spans approximately two and four orders of magnitude, respectively, and their values are not evenly distributed (Table S2). The log transformation compresses the dynamic range, reduces the influence of extreme values, and yields a distribution closer to normality, thereby facilitating model fitting and interpretation.

The scenario outlined above, further enables us to assess how the use of uncorrected versus corrected permeability values impacts predictive performance and, critically, the physicochemical interpretability of the resulting QSAR models. Importantly, the analysis and discussion presented below are intended in a constructive spirit—aimed at contributing additional evidence to the ongoing effort to refine experimental permeability measurements in narrow water channels such as AQ(G)Ps—rather than to endorse or disparage any particular experimental or analytical approach.

### 2.3. Pore Descriptors

Different computational resources were used to obtain or derive pore features (hereafter 'pore descriptors', PDs). Geometric PDs (including averaged pore diameter, minimum pore diameter and regularity) and pore-lining amino acid composition were obtained using the PoreWalker server [99]. Per-residue hydropathy (Kyte–Doolittle index [100]) was computed in Discovery Studio [101]. The Kyte-Doolittle hydrophobicity scale ranges from −4.5, the most hydrophilic value for Arg, to 4.5, the most hydrophobic value for Ile, with Gly being the most amphipathic residue (−0.4). PDs providing an indirect proxy for local mobility/flexibility were derived from crystallographic atomic displacements (B-factors) extracted from the PDB files. Finally, H-bonding- and electrostatic/solvation-related PDs were calculated following procedures analogous to those described by Horner *et al.* in Ref. [57] and Ref. [58], respectively (see Fig. 3).

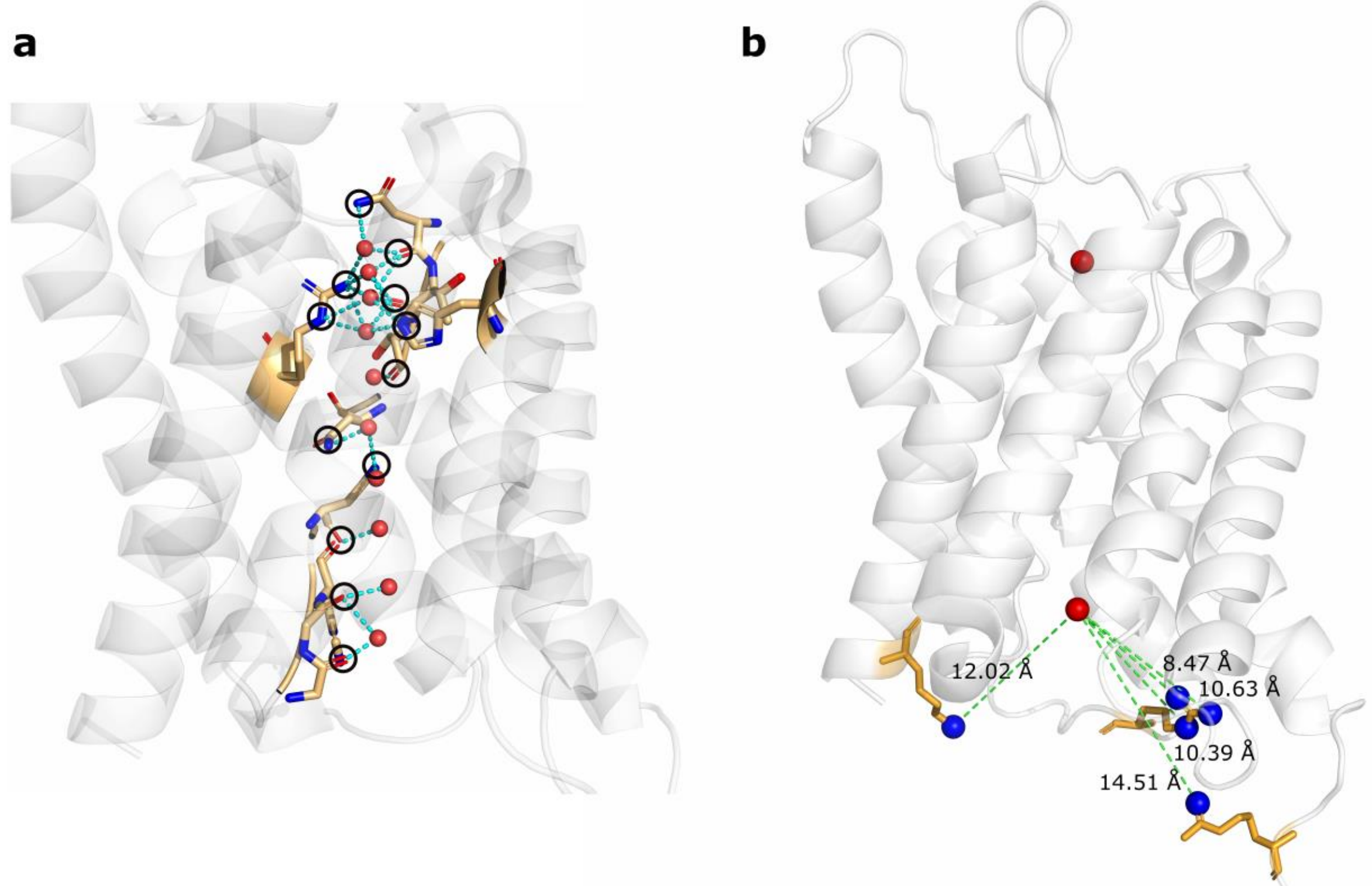

**Figure 3. Schematic illustration for eAqpZ (PDB ID: 1RC2) showing the computed number of potential pore-lining hydrogen-bonds ($N_H$), charged residues and charged sites located in the periplasmic and cytoplasmic pore vestibules.** (a) Calculated potential H-bonding sites ($N_H = 12$, black circles) along the single-file region

of eAqpZ. The single-file water column (red spheres) was defined following exactly the procedure described by Horner *et al*. [57]: water coordinates were taken from the 0.88 Å-resolution yeast AQP1 structure (PDB ID: 3ZOJ) [80] after structural superposition onto eAqpZ. Putative hydrogen bonds (cyan dashed lines) were identified using a 3.8 Å distance cut-off. (b) Positively charged Lys and Arg residues ($Res^{+} = 3$, tan-colored sticks), and associated charged side-chain nitrogen atoms ($Sites^{+} = 5$, blue spheres) counted within the pore cytoplasmic vestibule in eAqpZ, according to the established 15 Å cut-off to the reference waters (red spheres) represented on both the periplasmic and cytoplasmic sides. The red spheres mark the oxygen atoms of the first water molecules flanking the single-file water column, i.e., waters that are no longer part of the continuous single-file chain traversing the pore. Their coordinates were also taken from the above-mentioned yeast AQP1 structure (PDB ID 3ZOJ) [80] after structural superposition onto eAqpZ. The numbers next to the green dashed lines indicate the distances (Å)—all below the 15 Å cut-off (same defined by Horner et al. [58])—computed between each water oxygen atom (red spheres) and the charged side-chain nitrogen atoms (blue spheres) of the identified Lys and Arg residues within the same outer vestibule. Custom Python scripts were developed to compute these PDs and the rest of related pore features (see Table 2), including the number of negatively charged residues ($Res^{-}$, Glu and Asp) and charged sites/atoms ($Sites^{-}$) located in the pore vestibules.

When using the PoreWalker tool [99], the initial chain of each AQ(G)P structure was uploaded to the server for calculation. The PoreWalker output (a summary .TXT file is provided in the **Supplementary Information** for every AQ(G)P studied) includes the pore diameter profile along the pore *z*-coordinate, the 3D shape regularity percentage, 2D projection of the pore shape, cross-sectional images of the pore, as well as the list of amino acids lining the pore.

To carry out the present pore-level QSAR modeling, we first equalized the pore *z*-coordinate across the AQ(G)Ps with $p_f$ and/or $p_f^{corr}$ data compiled (Fig. 2 and Tables S1-S3). This was done by aligning all the pores along the *z*-axis, using as a common reference point the β-carbon (βC) of the asparagine residue in the second NPA motif. The βC coordinate was thus set to $z = 0$ (Table S3, Fig. 4a, and left panels in Fig. S1). After alignment, *z*-coordinates outside the range of −20 and 17.5 Å were not considered (Fig. 4f). This sectional "trimming" was motivated by the observation that pore diameter profiles for several AQ(G)Ps (obtained by PoreWalker [99]) showed artifactually elongated *z*-coordinates at either the intracellular or extracellular end, or both (see Fig. 4 and Fig. S1). In these regions, the pore shape is less well defined, making it difficult to unambiguously determine the residues lining the channel.

**Table 2.** Overview of the pore descriptors computed and analyzed in this work.

| Pore Descriptor (PD) | PD Short Name | PD Type | Description & Computation |
|---|---|---|---|
| Averaged diameter | $Ave_diam$ | Geometric | Diameter average computed from the diameter versus z-coordinate data provided by Pore-Walker server [99], along the pore z-coordinate-'trimming' analyzed. Computed by a custom script. |
| Minimum diameter | $Min_diam$ | Geometric | Minimum diameter identified from the diameter versus z-coordinate data provided by the Pore-Walker server [99]. |
| 3D shape regularity | $Reg$ | Geometric | Measure of how uniformly the pore geometry (cross-sectional symmetry) is maintained along the channel axis (z-coordinate). Reported by the PoreWalker server [99] as a percentage. |
| Averaged hydrophobicity indexes on pore-lining residues | $Ave_hydropho$ | Hydropathy | Average of per-residue hydrophobicity along the pore z-coordinate-'trimming' analyzed. Hydropathy indexes from the the Kyte-Doolittle scale [100], computed by Discovery Studio [98]. $Ave_hydropho$ accounts for strictly hydrophobic amino acids (i.e., those with positive hydrophobicity). |
| Averaged hydrophilicity indexed on pore-lining residues | $Ave_hydrophi$ | Hydropathy | Average of per-residue hydrophilicity along the pore z-coordinate-'trimming' analyzed. Hydropathy indexes from the Kyte-Doolittle scale [100], computed by Discovery Studio [98]. $Ave_hydrophi$ accounts for strictly hydrophilic amino acids (i.e., those with negative Kyte-Doolittle hydrophobicity). These negative indexes were converted into positive values—by multiplying by $-1$— to favor better interpretability. |
| Averaged per-residue B-factor on pore-lining residues within the selectivity filter region | $SF_Ave_Bfac$ | Mobility/flexibility | Average of per-residue displacement (B-factor) on those residues located within the pore *z*-coordinate 2-6 Å (see Fig. 4a and Fig. S1). B-factor values extracted from the PDB files using a custom script. |
| Average side chain B-factor on pore-lining residues within the selectivity filter region | $SF_Ave_Bfac_SC$ | Mobility/flexibility | Average of per-residue side-chain displacement on those residues located within the pore *z*-coordinate 2-6 Å (see Fig. 4a and Fig. S1). Extracted from the PDB file by a custom script. |

**Table 2.** Continuation…

| Pore Descriptor (PD) | PD Short Name | PD Type | Description & Computation |
|---|---|---|---|
| Averaged backbone B-factor on pore-lining residues within the selectivity filter region | $SF_Ave_Bfac_BB$ | Mobility/flexibility | Average of per-residue backbone displacement on those residues located within the pore *z*-coordinate 2-6 Å (see Fig. 4a and Fig. S1). Extracted from the PDB file by a custom script. |
| Averaged B-factor on pore-lining residues | $Ave_Bfac$ | Mobility/flexibility | Average of per-residue displacement on all the residues lining the pore. B-factor values extracted from the PDB files using a custom script. |
| Averaged side chain B-factor on the pore-lining residues | $Ave_Bfac_SC$ | Mobility/flexibility | Average of per-residue side chain displacement on all the residues lining the pore. B-factor values extracted from the PDB files using a custom script. |
| Averaged backbone B-factor on the pore-lining residues | $Ave_Bfac_BB$ | Mobility/flexibility | Average of per-residue backbone displacement on all the residues lining the pore. B-factor values extracted from the PDB files using a custom script. |
| Number of potential H-bonds with pore-lining residues along the single-file region | $N_H$ | H-bonding | Number of potential H-bonding sites along the single-file region. The single-file water column defined following exactly the procedure described by Horner *et al.* [93] (see the caption of Fig. 3, panel a). |
| Positively charged Lys and Arg residues in the pore vestibules | $Res^+$ | Electrostatic/solvation | Number of positively charged Lys and Arg residues counted within the pore outer vestibules (with side-chain nitrogen atoms within 15 Å distance cut-off to the reference waters in their own vestibule, see Fig. 3, panel b). It is a measure of enrichment of positively charged sites at the pore mouths, computed by a custom script. |
| Positively charged side-chain nitrogen atoms from Lys and Arg in the pore vestibules | $Sites^+$ | Electrostatic/solvation | Number of positively charged side-chain nitrogen atoms from Lys and Arg residues counted within the pore outer vestibules (within 15 Å distance cut-off to the above-mentioned reference waters, see Fig. 3, panel b). It is another measure of enrichment of positively charged sites at the pore mouths, computed by a custom script. |

**Table 2.** Continuation…

| Pore Descriptor (PD) | PD Short Name | PD Type | Description & Computation |
|---|---|---|---|
| Negatively charged Glu and Asp residues in the pore vestibules | $Res^{-}$ | Electrostatic/solvation | Number of negatively charged Asp and Glu residues counted within the pore outer vestibules (with side-chain oxygen atoms within 15 Å distance cut-off to the above-mentioned reference waters. It is a measure of enrichment of negatively charged sites at the pore mouths, computed by a custom script. |
| Positively charged side-chain nitrogen atoms from Lys and Arg in the pore vestibules | $Sites^{-}$ | Electrostatic/solvation | Number of negatively charged side-chain oxygen atoms from Asp and Glu residues counted within the pore outer vestibules (within 15 Å distance cut-off to the above-mentioned reference waters). It is another measure of enrichment of negatively charged sites at the pore mouths, computed by a custom script. |

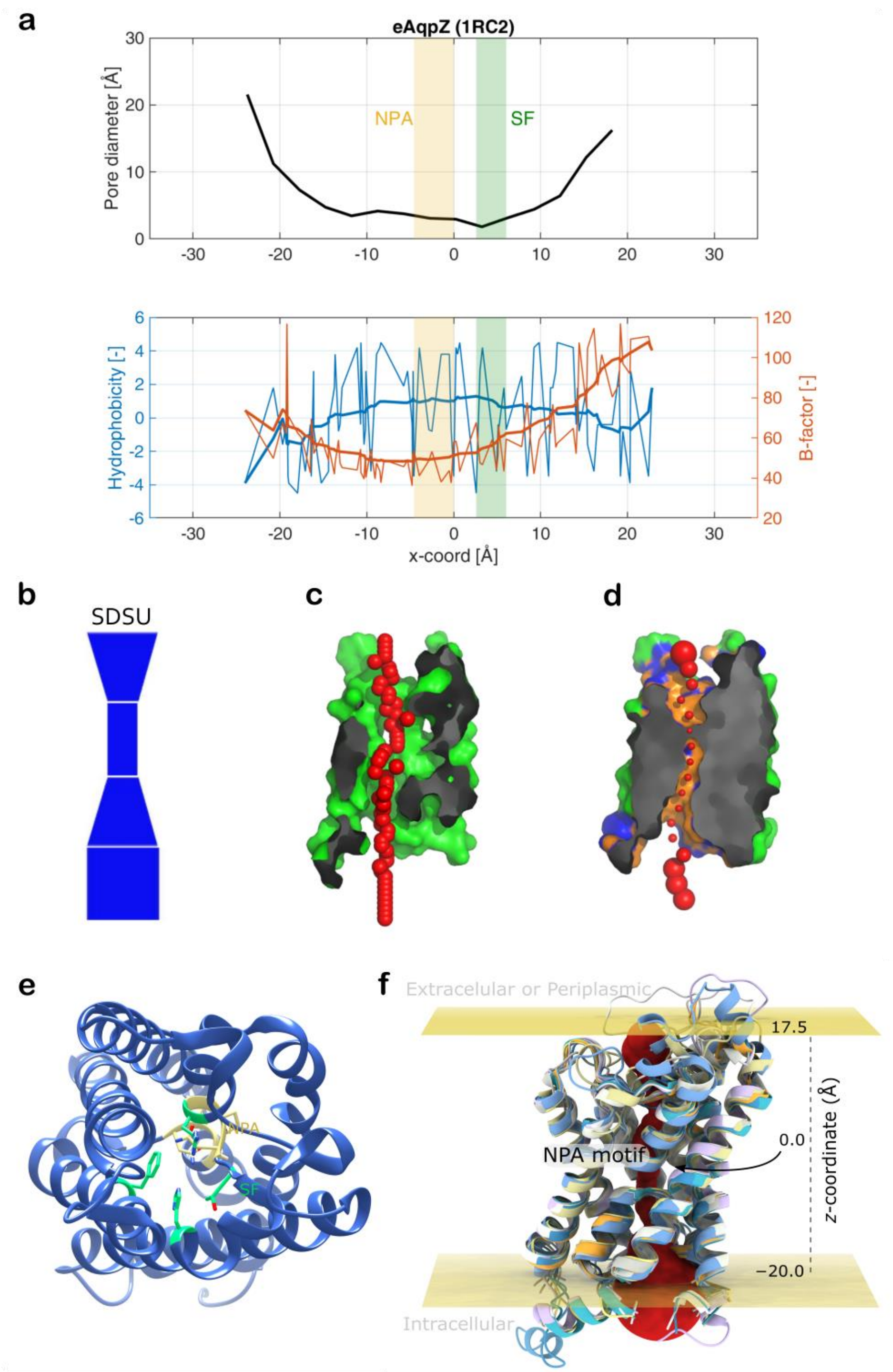

**Figure 4. Pore properties profiles and geometrical shape for *Escherichia coli* aquaporin Z (eAqpZ, PDB ID 1RC2), and scheme showing the *z*-coordinates equalization of the AQ(G)Ps modeled in this work**. (a) Top panel: diameter profile along the pore principal axis (*z*-coordinate) at 3Å step as calculated by the PoreWalker tool [99]; bottom panel: per-residue hydrophobicity index (blue) and B-factor (orange) of amino acids in the pore lining along the pore z-coordinate, as calculated by Discovery Studio (BIOVIA, Dassault Systèmes) [101] and as obtained from the crystal PDB file (1RC2), respectively. The thicker lines are smooth curves (noise removed). Vertical bands show the *z*-coordinates ranges where the NPA (orange) and SF (green) motifs are located. (b) Funnel-like form representing a 2D projection of the predicted pore shape as calculated by variations of diameter values along the pore: the bottom of the pore axis (i.e. the lowest *z*-coordinate) corresponds to the bottom of the stack. In the letter

code on the top of the form, 'D' indicates a conical frustum generated by decreasing pore diameter values, i.e. the diameter of the lower base of the frustum is bigger than the diameter of its upper circle, 'U' indicates the opposite conical frustum, i.e. lower base diameter smaller than upper base diameter, and 'S' indicates a cylinder (the code ordering from bottom to top). (c) and (d) Longitudinal cross-sectional images of the pore issued by PoreWalker, where the pore regularity is shown through red spheres located at the centre of the pore every 1Å step (c), whereas the pore dimension is represented by red spheres—at the center of the pore at 3Å step—whose sizes are proportional to the pore diameter calculated at that point (d). In the latter, orange and blue areas show pore-lining atoms and corresponding residues, respectively. (e) Cartoon representation of a single monomer of *E. coli* aquaporin Z (PDB ID: 1RC2), viewed from the extracellular side. The NPA motifs are shown as sticks in pale yellow (residues 63–65 for NPA-1 and 186–188 for NPA-2), and the selectivity filter residues (F43, H174, T183, and R189) are shown as green sticks. (f) Scheme showing the *z*-coordinates equalization performed on the modeled AQ(G)Ps. The central red yarn ball-like form approximately represents the pore diameter of one of the aligned AQ(G)Ps. **Notes:** Amino acids in the pore lining are reported in the PoreWalker's output files (**Supplementary Files**). The *z*-coordinates in the properties profiles (a) were shifted from the original ones issued by PoreWalker, to place the beta carbon (βC) of the asparagine (N) in the second NPA motif at the center of the *z*-coordinate system (*z*=0). This ensures that all the AQ(G)Ps are aligned to have the same reference *z*-coordinate, as showed in panel f.

Trimming the pore ends therefore helps to reduce the uncertainty in identifying the pore-lining residues and allows the application of a uniform framework across all the AQ(G)Ps, thereby enabling a more consistent calculation of pore properties, including average diameter, hydrophobicity, and B-factor–related pore descriptors.

All in all, thirteen (13) pore descriptors (PDs) were selected to conduct the pore-level QSAR modeling proposed in this study. Table 2 contains an overview of these PDs. A subset ($Reg$, $SF_Ave_Bfac$, $SF_Ave_Bfac_BB$ and $SF_Ave_Bfac_SC$) was first scaled to the 0–10 range. Then, standardization was applied to selected descriptors included in the QSAR models. These preprocessing steps were implemented to minimize scale-related bias during model training.

### 2.4 Pore-level QSAR Modeling and Performance Assessment

Given the limited number of AQ(G)Ps with compiled values of $p_f$ or $p_f^{corr}$ (n = 10 and n = 8, respectively) among those listed in Table 1, we use these entire sets to train the models. These two training datasets—containing the unitary log-transformed water permeabilities ($ln(p_f)$ or $ln(p_f^{corr})$) together with the calculated PDs—are provided as **Supplementary Files** in CSV format.

The iterative regularization technique LASSO (Least Absolute Shrinkage and Selection Operator) *[102]* was used, along with the Ordinary Least Squares (OLS) multiple linear regression method (MLR), to select optimal combinations of PDs for QSAR modeling. LASSO repeatedly fits a regression model by applying an L1 penalty so that small coefficients are shrunk to zero, thereby removing irrelevant variables and helping to prevent overfitting and multicollinearity.

Statistical parameters obtained to assess the performance of the QSAR models include: R-squared ($R^2$, internal coefficient of determination), regression *p*-value (from a Fisher F-test), statistical significance (from a Student t-test, α=0.05) for each independent variable (PD) in the models and the independent term, as well as the predictive parameters: SPRESS: Standard Prediction Error Sum of Squares, and loo-$Q^2$ (leave-one-out cross-validation coefficient).

Regarding the parameters mentioned above, $R^2$ is a measure of how well the model fits the data used to train the model (desirable $R^2$ is close to 1, whereas $R^2$ close to 0 denotes bad performance). The loo-$Q^2$ parameter is a measure of the internal predictive ability, and not just the fit. This metric is obtained after: (i) removing one item from the dataset, (ii) training again the model with the rest of items in the dataset, (iii) predicting the left-out item using the re-trained model. This process is then repeated until every item has been left out once. Then, all the prediction errors are combined into $Q^2$. Thus, higher loo-$Q^2$ (crucially, >0.5 up to 1) means that the model predicts well data internally, whereas low loo-$Q^2$ means it predicts poorly (basically noise).

To evaluate collinearity among the retained predictors, we inspected the pairwise correlation matrix and calculated variance inflation factors (VIFs) for each final descriptor set. Candidate models displaying pairwise intercorrelations $|r| > 0.5$ and/or VIF > 5 were excluded. In addition, the condition number (CN) was calculated to assess the sensitivity of the regression coefficients to perturbations in the data; models with CN > 5 were considered numerically unstable and were not pursued further (additional details are provided in the **Supplementary Information**).

A *y*-scrambling analysis was also conducted. The *y*-scrambling (also called *y*-randomization) is a method to assess chance correlation issues. In QSAR, an unscrambled model is the natural relationship that is observed between the dependent Y variable(s) Y and the independent X variable(s). A *y*-scrambled model, instead, is obtained by deliberately randomizing (shuffling) the Y values in the dataset while keeping X unchanged across the analyzed items. This procedure is done several times so that allows to verify that none of the "randomized" (scrambled) models perform better than the observed (trained) model, thus excluding the possibility that the proposed model is a result of accidentally fitting noise.

For model assessment, we took advantage of the availability of four aquaporins that, despite lacking experimentally resolved 3D structures, have $p_f^{corr}$ permeability values reported in the literature [56] (Table S2 and Fig. 2). These proteins were used as an external test set to evaluate one of the best-performing models developed in this study for the $p_f^{corr}$ dataset (see **Results and Discussion**). We used AlphaFold2-predicted 3D structures [103,104] for this purpose, and computed the pore descriptors included in the model in the same manner as for the rest of the dataset. Since AlphaFold2 does not provide B-factor information, and its confidence metrics (pLDDT and PAE) have been shown not to be equivalent to experimental B-factors [105,106], we restricted this validation only to the QSAR model that do not rely on B-factor features.

A dedicated Python script was developed to perform all regression, statistical, and *y*-scrambling analyses. This script is also supplied, together with the AQ(G)P PDB or AlphaFold structure files, as **Supplementary Data**. Further details on the pore-level QSAR modeling implemented in this study can be found in the **Supplementary Information** document.

## 3. RESULTS AND DISCUSSION

### 3.1. On AQ(G)Ps' Permeation Rate

Two key functional elements of AQ(G)Ps are their selectivity and permeation properties. In the **Supplementary Information**, we provide a systematic description of the main structural motifs that determine these properties, highlighting the similarities and differences across the representative set of AQ(G)Ps investigated in this study (see Table S4, Fig. S2 and Fig. S3). AQ(G)Ps' pore shape (also analyzed in detail in the **Supplementary Information**), whose conical design—both in the cytoplasmic and the extracellular entrances (Fig. 4b, Fig. 4f and Fig. S1)—ensures the creation of highly permeable water channels [60] that can transport ~$10^9$ water molecules per second (per monomer) in a single-file manner [107–110]. Both experimental [1,56,65,107,108,110–112] and in silico [8,63,98,109,113] approaches have been devised to measure or simulate the water permeation rate of AQ(G)Ps.

Access to reliable $p_f$ data of AQ(G)Ps is required to unravel the key structure–function relationships to a molecular level, as we intend here. Back in the early 90's, *in vitro* functional assays were developed to measure the osmotic water permeability of membrane water channels. Classical—though still used—swelling assays based either on the microinjection and expression of a transcribed protein sequence (typically the mRNA of the protein channel of interest) in *Xenopus laevis* oocytes (oocytes swelling assay: OSA) [1] or the reconstitution of the protein in liposomes (proteoliposome swelling assay: PSA) [107], have been indispensable tools for the estimation of the water permeability ($P_f$) and, by inference, unitary water permeabilities ($p_f$) for many AQ(G)Ps.

Despite the advances achieved in methodologies for measuring the water permeabilities of AQ(G)Ps [114]—particularly improvements in the physical techniques employed [65]—, the vast majority of unitary permeability measurements ($p_f$) have been obtained by indirect methods that heavily rely on the accurate estimation of the number of channels present in the sample. As stated by Gonzalez and collaborators, innovative methodological efforts are still needed to enable direct experimental measurements of the events occurring in water molecules inside AQ(G)P channels [51]. In the last decade, although still insufficient, some efforts have been made to obtain more direct permeation measurements [97] and to apply improved experimental strategies [115].

Notably, recent re-analyses have suggested that a non-negligible fraction of legacy stopped-flow–derived osmotic and unitary permeability estimates may be systematically overstated (up to 75%), owing to methodological assumptions that were common in earlier studies [56]. This has two practical implications for comparative structure–function analyses: first, absolute $p_f$ values across the literature should be interpreted with caution unless they derive from modern analysis pipelines or include explicit corrections; and second, relative rankings and broad trends are generally more robust than the original numerical magnitudes.

On the computational side, most approaches have relied on molecular dynamics simulations (MDS), including simulations under equilibrium conditions [63,98,116–123], pressure-induced [98], and steered molecular dynamics [113]. However, Zhu et al. [98] reported that equilibrium MDS is not a reliable method for computing osmotic unitary water permeabilities and suggested instead imposing a hydrostatic pressure difference across the membrane to better approximate experimental conditions [98]. Despite these limitations, MDS-based approaches—unlike experimental methods—enable the direct computation of $p_f$ under fully controlled conditions and have provided invaluable atom– and residue–level insights into filtration mechanisms, selectivity, and proton exclusion [50,63,85,98,117]. Moreover, they hold the potential to deepen our understanding of the behavior of protein channels in lipid membranes and to enable detailed thermodynamic and stability analyses [124,125]. Such analyses may help rationalize, and ultimately mitigate, the stability issues reported for several AQ(G)Ps [36].

### 3.2. Pore-level QSAR Study as a Methodological Proof of Concept

*3.2.1. Pore Descriptors (PDs)*

Several mechanistic studies have already attempted to quantitatively dissect how specific pore features modulate the unitary water permeability of narrow AQ(G)Ps, albeit typically by interrogating one factor at a time and/or using limited protein sets.

In particular, Horner *et al*. [57] proposed that single-file water mobility is governed by the number of potential hydrogen-bonds ($N_H$) between permeating waters and pore-lining residues, with transport limited by the kinetics of breaking and reforming these interactions during translocation. Closely related work further highlighted the role of entrance/exit dehydration–rehydration energetics, showing that the electrostatic environment at the pore mouth (e.g., enrichment of positively charged residues) can measurably boost $p_f$ even when the single-file H-bonding capacity is held constant, based on comparisons across a small panel of AQPs [58].

Complementary MDS-based analyses likewise indicated that permeability is not dictated by pore radius alone, but reflects a competition between channel geometry and the degree of single-file character along the permeation pathway [117]. Finally, experimental mutagenesis combined with in silico pore profiling has reinforced that subtle variations in constriction geometry and local physico-chemistry can differentially impact permeation [126], underscoring that “pore size” is rarely a sufficient standalone descriptor.

Notwithstanding these important advances—including the broader recognition that gating/regulatory phenomena can reversibly switch AQ(G)Ps between more and less permeable states under specific modulators (e.g., pH, ions, phosphorylation), adding another structural layer to consider—only a small number of pore descriptors have been rigorously benchmarked against extensive, harmonized experimental $p_f$ datasets to date. This motivates a dedicated descriptor framework for the pore-lining environment, in which geometric, H-bonding, electrostatic/solvation-related features, and other features are systematically encoded and evaluated jointly within a unified quantitative modelling context.

In the proposed pore-level QSAR modeling, we consider all pore features described above and additionally include descriptors that, to our knowledge, have not been systematically examined in the literature on determinants of water permeation, such as pore regularity and hydrophobicity- and B-factor–related pore descriptors (see Table 2).

Computed hydrophobicity profiles (Fig. 4a and Fig. S1) show significantly more hydrophobic residues (positive indexes) lining the pore than hydrophilic ones (negative indexes), in agreement with previous reports [127,128]. These differences are less pronounced near and at the intra- and extracellular pore vestibules, where the pore diameter and water contact surface increase, and there is a higher concentration of hydrophilic residues. The hydrophobicity profiles, thus, predominantly show a downward bell-shaped smoothed curve, with the peak located around the central part of the pore, where the NPA and SF motifs are situated. It is important to note that, although hydrophobic residues predominate—particularly in the central region of the pore—the presence of hydrophilic residues interspersed among them could be essential for water passage [62,117]. In relation to this, a pore lining composition analysis of the AQ(G)Ps listed in Table 1 is provided in the **Supplementary Information** (see Fig. S4) for the reader's reference.

Moreover, B-factor profiles display the opposite shape, that is, an upward bell smoothed curve (Fig. 4a and Fig. S1). The B-factor values, which are typically derived from fitting X-ray diffraction data, measure the atomic displacement in molecular structures, indicating how much atoms deviate from their average positions due to thermal motion or structural disorder. According to these graphs, residues in the pore with the lowest B-factors are located in the region encompassing the pore constriction sites, namely the SF and NPA motifs. This is likely due to the proximity of the pore walls, which facilitates the formation of interaction networks that are crucial for the transport and selectivity exhibited by these proteins. Accordingly, within our QSAR framework, we evaluated B-factor–derived descriptors at the whole-residue, backbone, and side-chain levels, computed as averages over residues in the selectivity filter region as well as over the all pore-lining residues for the AQ(G)Ps analyzed.

We additionally computed B-factor descriptors at the whole-protein level and for the extra-pore region (i.e., non–pore-lining residues) using the same structure alignment and pore-trimming workflow. However, none of these descriptors was statistically significant and, accordingly, they provided no informative signal in our tests. These descriptors are therefore not included in Table 2, which focuses on pore-related descriptors. Instead, their computed values are provided in the **Supplementary Information** as a separate file for transparency.

#### *3.2.2. Pore-level QSAR Models: Performance and Mechanistic Insights*

We present here the best-performing models ($R^2 > 0.9$ and loo-$Q^2 > 0.6$) obtained for the two permeability datasets, following the criteria defined in Section 2.4: (1) AQP(G)Ps with compiled $p_f$ (n=10), and

(2) AQP(G)Ps with compiled $p_f^{corr}$ (n=8). Candidate QSAR models containing four or more PDs exhibited substantial multicollinearity and/or instability and were therefore discarded, even when their apparent goodness of fit was high (data not shown).

***Model 1.3*** *(uncorrected permeabilities; n = 10):*

$$ln(p_f) = -1.38(\pm 0.21) * N_H + 0.98(\pm 0.20) * SF_Ave_Bfac_SC - 1.18(\pm 0.22) * Sites^- + 0.83(\pm 0.19)$$

This model exhibits an excellent fit ($R^2$ = 0.92; Adj. $R^2$ = 0.88; Prob(F) = 0.001) together with moderate-to-good internal predictivity (loo-$Q^2$ = 0.71). In real-scale terms, predicted $p_f$ values remained within the same order of magnitude as the experimental (observed) values across all cases. The largest absolute deviation (12.5 x $10^{-14}$ $cm^3 \cdot s^{-1}$) was observed for the highest permeability channel (rat AQP4; Table 4 and Fig. 5a), which is an expected feature of small-n QSAR models where high-leverage points can dominate.

The strongest effect in Model 1.3 is the negative coefficient for $N_H$, defined as the number of potential H-bonding sites along the single-file region. This indicates that increasing the density of H-bond-capable pore-lining residues is associated with reduced water permeability. The direction of this effect is consistent with the single-file transport framework, in which wall–water H-bonding introduces kinetic friction: permeating waters must repeatedly break and reform interactions with pore-lining residues, slowing collective motion through the constriction. The dependence on $N_H$ in Model 1.3 is also consistent with prior work by Horner *et al.* [57], who reported an inverse relationship between $p_f$ and $N_H$ across a small set of AQ(G)Ps, protein channels, and nanotubes.

The positive coefficient for $SF_Ave_Bfac_SC$ (average side-chain displacement/B-factor for residues within *z* = 2–6, i.e., around the selectivity filter) suggests that greater local side-chain mobility in the constriction region correlates with increased water permeability in this dataset. A plausible mechanistic interpretation is that moderate flexibility could transiently relax steric or interaction constraints, smoothing the free-energy landscape for single-file water translation (e.g., by reducing the persistence of “sticky” configurations). However, because B-factors conflate genuine dynamics with crystal packing, resolution, and refinement procedures, this descriptor should be interpreted cautiously as a proxy for flexibility rather than a direct dynamical observable.

The negative coefficient for $Sites^-$ indicates that a higher density of negatively charged sites in the outer vestibules is associated with reduced $p_f$ in this dataset. This observation is compatible with, but also extends, previous mechanistic analyses of aquaporin entrances. In particular, Horner *et al.* reported a clear positive association between positively charged residues at the channel mouth and single-file water flow, whereas the contribution of negatively charged residues appeared comparatively marginal in the small set of channels examined [58]. In Model 1.3, however, the LASSO procedure retained $Sites^{-}$—defined not as the number of negatively charged residues, but as the number of negatively charged oxygen atoms from Asp/Glu side chains within a defined vestibular proximity to reference waters (see Table 2)—suggesting that a more fine-grained representation of vestibular anionic functionality may capture effects that are not resolved by residue counts alone.

A plausible interpretation is that vestibular anionic groups can oppose water entry and throughput by increasing the energetic cost of reorganizing hydration upon approach to the pore mouth, thereby counteracting the permeability-enhancing effect of cationic sites (with a lower desolvation penalty, see discussion below for Model 2.3b) [58]. In addition, an increased abundance of Asp/Glu oxygen atoms provides additional opportunities for transient water–protein hydrogen bonding in the vestibular regions, which could slow water exchange into the single-file pathway by raising the local residence time and effectively increasing friction before and/or after the principal constriction. These mechanisms are not mutually exclusive and may act in concert. Given the limited sample size (n = 10), the contribution of $Sites^{-}$ is best interpreted as a mechanistically plausible association that merits targeted validation, rather than as a definitive causal determinant of $p_f$.

***Model 2.3a*** *(corrected permeabilities; n=8)*:

$$ln(p_f^{corr}) = -0.53(\pm 0.21) * SF_Ave_Bfac_SC - 0.77(\pm 0.21) * Reg - 1.55(\pm 0.21) * N_H + 1.57(\pm 0.20)$$

This model shows an excellent goodness-of-fit ($R^2$ = 0.95; Adj. $R^2$ = 0.92; Prob(F) = 0.004) but a more modest internal predictability (loo-$Q^2$ = 0.63). In real-scale terms, predicted $p_f^{corr}$ values remained within the same order of magnitude as the compiled (observed) permeabilities across all AQ(G)Ps. The largest absolute deviation was observed for *E. coli* GlpF (26.8 x $10^{-14}$ $cm^3 \cdot s^{-1}$; Table 4 and Fig. 5b).

Notably, $N_H$ remains a dominant negative term, reinforcing the robustness of the H-bond friction concept across datasets. The negative coefficient for $Reg$ (PoreWalker-derived regularity [100]) suggests that a more uniform pore geometry along the pore axis is associated with lower permeability in the corrected dataset. At this stage, $Reg$ is best viewed as a global structural correlate of permeability rather than a direct mechanistic determinant. Clarifying its physical meaning will likely require a more detailed analysis that resolves local radius variations along the pore and quantifies how these features couple to the free-energy barriers governing single-file water flow.

In contrast to Model 1.3, the coefficient associated with $SF_Ave_Bfac_SC$ is negative in Model 2.3a. However, this term also displays the smallest effect size among the selected predictors (−0.53 in standardized units) and the weakest statistical support within the model (P>|t| = 0.064, whereas the remaining terms show P-values ≤ 0.01). Taken together, the modest magnitude, marginal significance, and sign reversal relative to the non-corrected-permeability model suggest that B-factor–derived descriptors could be particularly sensitive to changes in the response definition (i.e., corrected versus non-corrected permeabilities) and to dataset composition. In addition, the training set comprises structures obtained under heterogeneous experimental conditions (including resolution, temperature, pH, crystallization/buffer conditions, and structure-determination method; see Table S6), all of which can influence crystallographic displacement parameters and limit the direct comparability of B-factors across AQ(G)Ps. Accordingly, at this stage, the $SF_Ave_Bfac_SC$ term should be interpreted cautiously and revisited once a larger dataset becomes available.

***Model 2.3b*** *(corrected permeabilities; n=8)*:

$$ln\left(p_f^{corr}\right) = 0.39(\pm 0.39) * Sites^{+} - 0.95(\pm 0.33) * Reg - 1.59(\pm 0.35) * N_H + 1.57(\pm 0.29)$$

This model also provides a strong fit ($R^2$ = 0.90; Adj. $R^2$ = 0.826; Prob(F) = 0.018) together with similar moderate internal predictivity (loo-$Q^2$ = 0.65). In real-scale terms, predicted $p_f^{corr}$ values remained within the same order of magnitude as the compiled (observed) permeabilities for all channels, with the exception of *E. coli* GlpF, for which the absolute discrepancy reached 46.6 x $10^{-14}$ $cm^3 \cdot s^{-1}$ (Table 4 and Fig. 5c). Relative to Model 2.3a, Model 2.3b replaces the $SF_Ave_Bfac_SC$ descriptor with $Sites^{+}$, defined as the number of positively charged Lys/Arg side-chain nitrogen atoms within the pore vestibules (Table 2). The positive coefficient for $Sites^{+}$ is consistent with previous evidence that positive charges at the pore mouth (entrances/exits) can boost water flow [58]. Accordingly, Model 2.3b supports the interpretation that vestibular cationic groups may facilitate water entry and/or reduce the energetic penalty associated with reorganisation of hydration at the pore mouth, thereby promoting the initiation and maintenance of the single-file water column.

Importantly, Model 2.3b is the most amenable to external testing among the QSAR models presented here, as it relies on descriptors that can be computed from AlphaFold2 [103,104] structures (in contrast to B-factor–based terms). External evaluation on four AQPs (rAQP8, mAQP11, RsAqpZ and sAQP) shows good overall agreement between predicted and compiled $p_f^{corr}$ values (Fig. 5d), with $R_{Ext}^2 = 0.898$. In real-scale terms, absolute deviations ranged from 0.85 to 8.77 × $10^{-14}$ $cm^3 \cdot s^{-1}$; in relative terms, three AQPs were predicted within ~2-fold (rAQP8, mAQP11 and RsAqpZ), whereas the lowest-permeability case (sAQP) was overestimated (~6.5-fold). Given the very small external set (n = 4) and the wide range in $p_f^{corr}$ values across this subset of AQPs, these results should be interpreted as preliminary evidence of transferability rather than as a definitive validation.

Overall, $N_H$ displays a consistent negative association with permeability across all models, supporting the view that an increased density of potential wall–water hydrogen-bonding sites in the single-file region imposes an effective friction that limits water throughput. By contrast, descriptors derived from crystallographic B-factors show dataset-dependent behavior, including a reversal in coefficient sign between the non-corrected and corrected permeability datasets, reinforcing their interpretation as empirical proxies rather than robust mechanistic determinants. Importantly, vestibular electrostatics emerge as a complementary modulator of permeability: the positive association of $Sites^+$ with $p_f^{corr}$, together with the negative trend observed for $Sites^-$ in the non-corrected dataset, is consistent with the notion that charge distribution at the pore mouth can influence water entry/exit by altering hydration reorganization and the associated desolvation penalty.

From a statistical standpoint, multicollinearity was negligible (VIF ≈ 1–1.76; condition number ≤ 2.21) across all models, supporting coefficient stability and interpretability. Moreover, *y*-scrambling analysis (100 permutations; Fig. S5, panels b, d and f) yielded no scrambled model with superior $R^2$/loo-$Q^2$ compared with the corresponding original model, which argues against chance correlations being responsible for the reported fits. All statistical parameters for the QSAR models are summarized in Table 3, and the per-descriptor LASSO coefficients resulted after the application of the feature-selection procedure are shown in the bar plots in Fig.S5 (panels a, c and e).

**Table 3.** Best-performing pore-level QSAR models and their associated statistical fit and predictive performance metrics.

| Training Dataset[a] | No. PDs in model[c] | Model ID | PD1 (coeff. ± SE)[d] | PD2 (coeff. ± SE)[d] | PD3 (coeff. ± SE)[d] | Indep. term (± SE)[d] | $R^2$ | P-value (F-test) | CN[e] | VIF[f] | loo-$Q^2$[g] | SPRESS[h] |
|---|---|---|---|---|---|---|---|---|---|---|---|---|
| 1 $ln(\mathrm{p_f})$ | 3 | 1.3 | $N_H$ (−1.38±0.21)*** | $SF_Ave_Bfac_SC$ (0.98±0.20)*** | $Sites^-$ (−1.18±0.22)*** | 0.84 ± 0.19*** | 0.92 | 0.001 | 1.68 | 1.20; 1.08; 1.29 | 0.71 | 0.87 |
| 2 $ln(\mathrm{p_f^{corr}})$ | 3 | 2.3a | $Ave_Bfac_SC$ (−0.53±0.21)* | $Reg$ (−0.77±0.21)** | $N_H$ (−1.55±0.21)*** | 1.57 ± 0.20*** | 0.95 | 0.004 | 1.38 | 1.07; 1.03; 1.11 | 0.63 | 1.12 |
| | | 2.3b | $Sites^+$ (0.39±0.38)* | $Reg$ (−0.95±0.33)** | $N_H$ (−1.59±0.35)*** | 1.57 ± 0.29*** | 0.90 | 0.018 | 2.21 | 1.30; 1.41; 1.76 | 0.65 | 1.09 |

[a]Water permeability datasets used to train the QSAR models presented in the rows: 1) dataset of individual or averaged $p_f$ measurements—without correction—for: bAQP0, oAQP0, bAQP1, hAQP1, hAQP2, hAQP4, rAQP4, mAqpM, eAqpZ and eGlpF (as reported in Table S1 and depicted in Fig. 2); 2) dataset of corrected water permeabilities for: oAQP0, bAQP1, hAQP1, hAQP2, hAQP4, rAQP4, eAqpZ and eGlpF (as reported in Table S2 and depicted in Fig. 2), [c]Number of PDs included in the QSAR models presented in the rows. [d]S.E. stands for standard error, and the number of asterisks shown over the values indicates the statistical significance (p-value from a Student t-test) obtained for the PDs or the independent term of the models, namely: * if $p < 0.1$, ** if $p < 0.05$, and *** if $p < 0.01$. [e]The condition number (CN) obtained for each model. A summarized description of CN is given in the **Supplementary Information**. [f]The Variance Inflation Factor (VIF) for all the variables (PDs) in the models. A summarized description of VIF is also provided in the **Supplementary Information**. [g]loo-$Q^2$ is the leave-one-out cross-validation coefficient used to assess the internal predictability of the models. [h]SPRESS stands for Standard Prediction Error Sum of Squares.

304

**Table 4.** Measured or corrected water permeation rates (observed), predicted values, and residuals obtained from Models 1.3, 2.3a and 2.3b in Table 3.[a]

| AQ(G)P (PDB) | Observed $p_f$ ($\times\ 10^{-14}$ $cm^3 \cdot s^{-1}$)[b] | Predicted $p_f$ ($\times\ 10^{-14}$ $cm^3 \cdot s^{-1}$) | Residual $p_f$ (Pred. minus Obs.) | Observed $p_f^{corr}$ ($\times\ 10^{-14}$ $cm^3 \cdot s^{-1}$)[c] | Predicted $p_f^{corr}$ ($\times\ 10^{-14}$ $cm^3 \cdot s^{-1}$) | Residual $p_f^{corr}$ (Pred. minus Obs.) | Observed $p_f^{corr}$ ($\times\ 10^{-14}$ $cm^3 \cdot s^{-1}$)[c] | Predicted $p_f^{corr}$ ($\times\ 10^{-14}$ $cm^3 \cdot s^{-1}$) | Residual $p_f^{corr}$ (Pred. minus Obs.) |
|---|---|---|---|---|---|---|---|---|---|
| | | **Model 1.3** | | | **Model 2.3a** | | | **Model 2.3b** | |
| eGlpF (1FX8) | 1.35 | 2.71 | 1.36 | 110.0 | 83.2 | −26.8 | 110.0 | 63.4 | −46.6 |
| bAQP1 (1J4N) | 2.3 | 2.95 | 0.65 | 3.2 | 2.93 | −0.27 | 3.2 | 2.75 | −0.45 |
| eAqpZ (1RC2) | 6.0 | 4.91 | −1.09 | 7.33 | 6.75 | −0.58 | 7.33 | 7.95 | 0.62 |
| bAQP0 (1YMG) | 0.15 | 0.25 | 0.1 | -- | -- | -- | -- | -- | -- |
| oAQP0 (2B6O) | 0.2 | 0.21 | 0.01 | 0.09 | 0.11 | 0.02 | 0.09 | 0.11 | 0.02 |
| rAQP4 (2D57) | 23.0 | 35.5 | 12.5 | 7.36 | 17.94 | 10.58 | 7.36 | 24.52 | 17.16 |
| mAqpM (2F2B) | 0.65 | 0.46 | −0.19 | -- | -- | -- | -- | -- | -- |
| hAQP4 (3GD8) | 11.0 | 10.0 | −1.0 | 11.0 | 11.5 | 0.5 | 11.0 | 4.45 | −6.55 |
| hAQP2 (4NEF) | 9.3 | 6.0 | −3.3 | 3.27 | 1.89 | −1.38 | 3.27 | 4.17 | 0.9 |
| hAQP1 (7UZE) | 5.1 | 2.12 | −2.98 | 4.6 | 3.77 | −0.83 | 4.6 | 4.12 | −0.48 |

[a]All unitary permeation rates and their associated residuals are presented on the original (non-log) scale, obtained by back-transforming the modeled log-transformed data. [b]Experimental water permeation rates (averages if more than one is reported for a given AQ(G)P), as listed in Table S1 (original $p_f$ values, i.e., without the correction proposed by Wachlmayr *et al.* [56]) . [c]Corrected water permeation rates [56] compiled in Table S2 and Fig. 2.

305

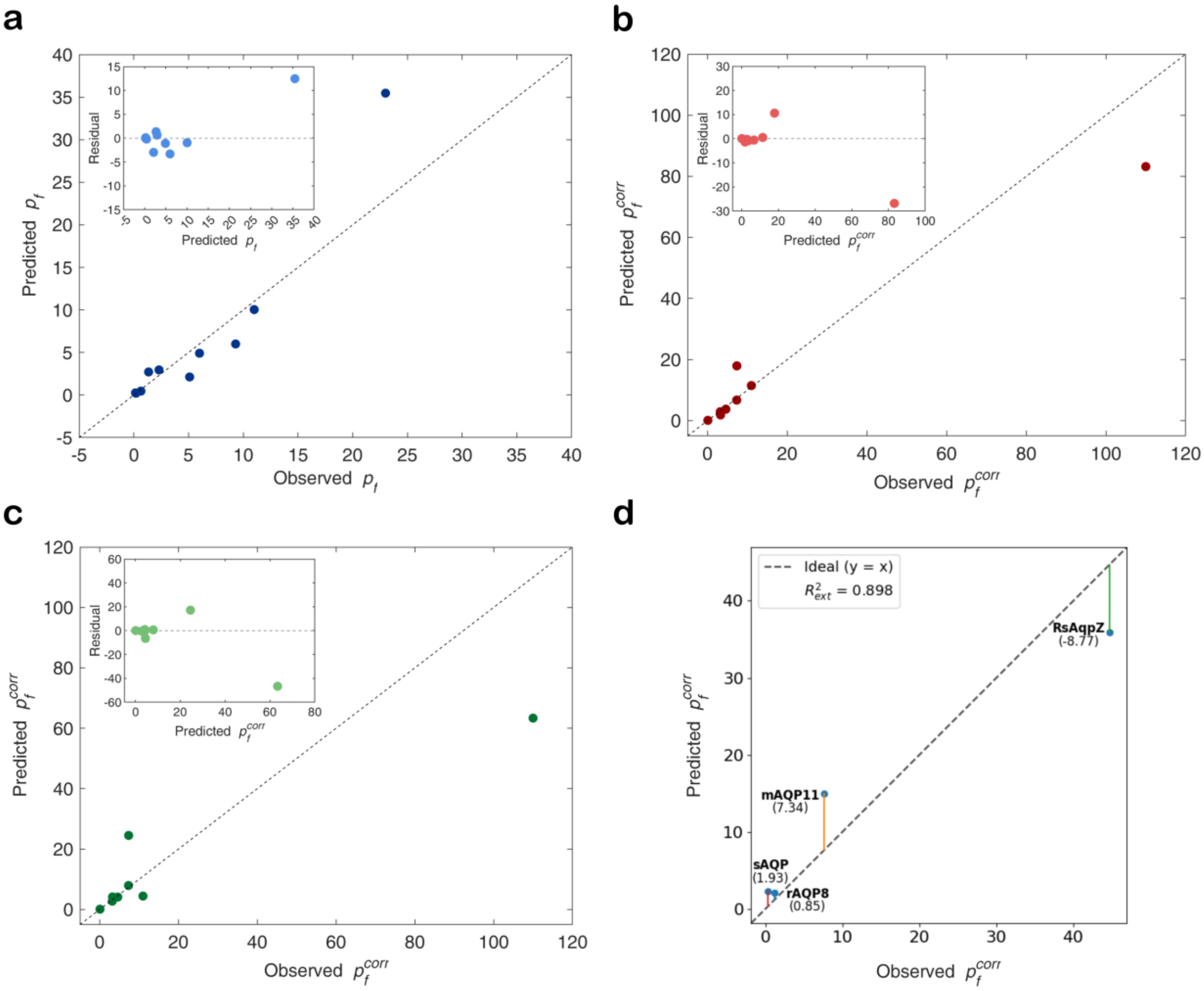

**Figure 5. Predictive performance and evaluation of the proposed pore-level QSAR models.** (a) Predicted versus observed non-corrected permeability ($p_f$) values obtained from Model 1.3 (Table 4). (b) Predicted versus observed corrected permeability ($p_f^{corr}$) values obtained from Model 2.3a (Table 4). (c) Predicted versus observed corrected permeability ($p_f^{corr}$) values obtained from Model 2.3b (Table 4). In (a-c) the diagonal dashed lines show the 1:1 relationship, and the insets display residuals versus predicted $p_f^{corr}$. (d) Predicted versus observed $p_f^{corr}$ values obtained with Model 2.3b for an independent set (rAQP8, mAQP11, RsAqpZ and sAQP; Table S6) used for external validation. The diagonal dashed line corresponds to the ideal prediction (*y* = *x*). The external coefficient of determination ($R^2_{Ext} = 0.898$) indicates a strong overall agreement between observed and predicted values. For each data point, the deviation from the diagonal (Δ = predicted – observed) is indicated in parentheses, highlighting over– and under–predictions by the model.

### *3.2.3. Challenges and Main Limitations of the Proposed Methodology*

Modelling water permeation through narrow channels such as AQPs and AQGPs remains intrinsically challenging. This complexity stems, on the one hand, from the coupling between local and global phenomena occurring both within and around the pore, including auto-regulatory processes (e.g., pore gating; discussed in more detail in the **Supplementary Information**, Fig. S6) and post-translational modifications [13,87,129–132]. On the other hand, defining descriptors that faithfully capture the structural and physicochemical determinants of permeation is not straightforward a priori.

In addition, the still limited availability of experimentally determined osmotic permeability coefficients ($p_f$) across AQ(G)Ps precludes the development of more robust predictive models based on extended training sets and, at the same time, restricts the possibility of reserving independent external test sets for rigorous validation.

Beyond data availability, substantial room remains to devise and validate new pore features / descriptors that encode the physical determinants of pore-mediated filtration and selectivity, an area where computational approaches could play a particularly important role. In this context, MDS-based methodologies could be leveraged to derive dynamic descriptors that go beyond static structural proxies. This perspective is increasingly feasible given the availability of high-quality structure predictors such as AlphaFold [103,104], together with growing access to high-performance computing resources. In particular, AlphaFold-derived structures of AQ(G)Ps can enable descriptor calculation and, consequently, support broader model development and validation—as illustrated here by the external set used to validate Model 2.3b (Table S6)—particularly as larger datasets of AQ(G)Ps with experimentally measured osmotic permeability become available.

It should also be noted that crystal structures, such as those used here, are not fully representative of proteins embedded in a lipid membrane. However, because our modeling framework focuses on the pore region, it is reasonable to assume that pore geometry and key constriction features are comparatively less sensitive to the surrounding membrane environment than other protein regions.

For this study, AQ(G)P structures solved at the highest available resolution were selected. In addition, to avoid biases and minimize inter-structure differences, all B-factor-based descriptors were computed after structural alignment of all AQ(G)Ps and applying identical pore segment definitions (z-coordinate pore trimming in the range of −20 to 17.5 Å). Nonetheless, modest differences remain across structures in terms of resolution, crystallization conditions (e.g., temperature, pH, and solvent), and data-collection protocols across the AQ(G)Ps analyzed here (Table S5). This issue limits direct comparisons between AQ(G)Ps and, particularly for models relying on B-factor features, complicate interpretation in terms of “true” dynamics or flexibility.

Despite these limitations, the good statistical metrics and predictive performance achieved by the selected QSAR models, together with their consistency with previously reported experimental observations (see previous section), render the proposed approach particularly encouraging. Importantly, this framework enables the joint assessment of multiple pore features and their combined effects on permeability. In our view, the results presented here may also motivate continued improvements in experimental procedures for determining osmotic water permeability, as well as the development of richer and more physically grounded pore descriptors, ultimately extending the applicability and robustness of this QSAR framework in future studies. Nonetheless, experimental validation and prospective design tests will be required before any predictive or engineering capability can be firmly established.

## 3. CONCLUSIONS

A pore-level QSAR approach is proposed to help identify key characteristics influencing water permeation through molecular pores. The best QSAR models achieved, while showing good correlation and moderate-to-good internal predictive performance, still present some limitations due to the scarcity of experimental permeation rate data for AQ(G)Ps, which hinders the use of broader training sets. Nevertheless, this constraint does not diminish the potential of the approach for gaining a deeper residue-level understanding of AQ(G)Ps' filtering properties, particularly by integrating multiple pore descriptors within a single quantitative framework rather than interrogating them individually. We believe that this methodology holds significant promise and, as additional permeation data become available, could yield a more comprehensive view of the pore permeation process. Ultimately, it may also support the rational design of highly efficient AQ(G)Ps or other biomimetic molecules with tailored filtration properties and provide further insights into the molecular mechanisms underlying AQ(G)P-related diseases.

## SUPPLEMENTARY INFORMATION

This article has an associated **Supplementary Information** document and **Supplementary Files**.

## CRediT authorship contribution statement

**J.J.G-F.**: Conceptualization, Methodology, Software, Validation, Formal analysis, Investigation, Data curation, Writing - Original draft preparation, Writing - Reviewing and Editing, Visualization, Project ad-ministration; **L.B.**: Conceptualization, Methodology, Software, Validation, Formal analysis, Investigation, Data curation, Writing - Original draft preparation, Writing - Reviewing and Editing, Visualization, Project administration; **P.V.**: Investigation, Data curation, Writing - Original draft preparation; **M.M.**: Writing - Reviewing and Editing; **M.F.**: Writing - Reviewing and Editing; **D.P.**: Writing - Reviewing and Editing; **E.C.**: Writing - Reviewing and Editing; **M.C.dR.**: Writing - Reviewing and Editing, Supervision.

## Data Availability Statement

Files with the raw and processed data, and the custom Python script prepared for the pore-level QSAR modeling can be accessed from: https://zenodo.org/records/18675157.

## Competing Interests

The authors declare no competing interests.

# Supplementary Information

**for**

*Pore-level Quantitative Structure-Activity Relationship (QSAR) for Water Permeation Rate in Aquaporins*

*Juan José Galano-Frutos[1,a,†,*], Luca Bergamasco[2,a,*], Paolo Vigo[2], Matteo Morciano[2], Matteo Fasano[2], Davide Pirolli[1], Eliodoro Chiavazzo[2] and Maria Cristina de Rosa[1]*

*[1] Istituto di Scienze e Tecnologie Chimiche "Giulio Natta" (SCITEC) - National Research Council (CNR), Via Largo Francesco Vito 1, 00168 Rome, Italy*

*[2] Department of Energy, Politecnico di Torino, C.so Duca degli Abruzzi 24, 10129 Turin, Italy*

*[†] Present address: Certest Biotec S.L., 50840 San Mateo de Gallego, Zaragoza, Spain*

*[a] These authors contributed equally to this work*

*[*] Corresponding author(s): J.J.G.-F. jjgalano@certest.es; L.B. luca.bergamasco@polito.it*

## Best Practices and Pore-level QSAR Modeling

*Multicollinearity analysis*. Auto-correlation matrixes were computed to identify multicollinearity among descriptors. In addition, the variance inflation factors (VIF)[1] and the condition number (CN)[2]—calculated for the best QSAR models presented in the main text (Table 3)—were used as complementary indicators of multicollinearity between variables.

*LASSO iterative regularization* [3]. LASSO was applied to the OLS-MLR models to select optimal combinations of PDs. The associated alpha[3] parameter (lambda in Ref. [3]) was set in the range 0.1–1.0 (moderate penalty) to discourage highly correlated predictors. The maximum number of iterations was set to 1,000,000, and convergence was achieved in all cases.

## Characteristic Features of AQ(G)Ps: Similarities and Differences

<u>*Basic Structure*</u>. Aquaporins are tetramers formed by four monomer channels, each constituted by six transmembrane and two half-spanning helices interconnected through five loops [4–6]. The inner terminals of the two half helices host the amino acid sequences constituting the NPA motifs, which face each other in the middle of the pore (Fig. 1 of the main document). This motif typically involves asparagine, proline and alanine (Asn, Pro and Ala) residues, which are highly conserved in aquaporins [5–24] (see the conservation analysis addressed below). In this region, proton exclusion is accomplished with a reorientation of the water molecules, which favors their passage [4, 7, 25, 26].

---

[1] The Variance Inflation Factor (VIF) [1] measures the extent of multi-collinearity in a regression model. A VIF value of 1 indicates no correlation between the independent variable and others. Values between 1 and 5 suggest moderate multicollinearity, while values above 5 indicate significant multicollinearity that could affect the stability of coefficient estimates. A VIF greater than 10 is often considered problematic, as it implies substantial inflation of the variance of the estimated coefficients, leading to unreliable results.

[2] The condition number (CN) [2] is a crucial metric that assesses the sensitivity of the regression coefficients to perturbations in the data. Generally, a condition number less than 10 indicates a well-conditioned model, suggesting stable and reliable coefficient estimates with minimal numerical errors. Values between 10 and 30 are often associated with moderate multicollinearity, where the results may still be usable but should be approached with caution. Condition numbers between 30 and 100 indicate significant multi-collinearity, raising concerns about the stability of the coefficient estimates. A condition number greater than 100 signals a poorly conditioned model, which can lead to high sensitivity to input changes, resulting in considerable numerical instability and unreliable outputs.

[3] The alpha parameter in the LASSO method controls the strength of the regularization applied to the regression model. A small alpha value (e.g., <0.1) results in minimal regularization, allowing for more complex models that may overfit the data. Conversely, a larger alpha value (e.g., >1.0) increases the penalty on the coefficients, particularly affecting variables with significant multicollinearity. This leads to simpler models with potentially better generalization, but at the cost of possibly excluding important predictors and risking underfitting the data.

Each monomer presents a second-stage filtering feature in its channel that is located closer to the extracellular border [5, 25] and works as a constriction region. This feature is generally referred to as selectivity filter (SF) or ar/R (aromatic/arginine) region, since it always comprises at least one aromatic residue and one Arg. On the outside of the pore, helices are hydrophobic, while on the inside they are mainly hydrophilic [27]. Each monomer is functionally independent with its own pore, and when assembled together, the four monomers form a pore in the middle (Fig. 1a of the main text). No water permeation has been observed in natural AQ(G)Ps through this central pore, so that the flux takes place only through the four lateral pores of the monomer units *[28]*. While the orthodox AQPs' pore has been shown to have approximately 1-3 Å minimum diameter [26], AQGPs exhibit a constriction pore passage which is approximately 1 Å wider, which allows them to accomplish glycerol diffusion [27]. In the following sub-sections, we will present a detailed analysis of the two main features described above (see Table S4), along with an examination of the pore diameter and geometric characteristics across the selected set of AQ(G)Ps. In addition, we include a systematic analysis where we conducted structure- and sequence-based comparisons, to highlight the similarities and differences exhibited by these fundamental filtering molecules (see Fig. S2 and Fig. S3).

*NPA Motif.* The AQ(G)P family features two highly conserved asparagine-proline-alanine (NPA) motifs that are absent in other protein channels, like gramicidin A [7]. These two motifs, located in the narrow central segment of the channel are essential for water permeation and selectivity. Proton exclusion in AQ(G)Ps, which is essential to maintain the electrochemical gradient across cellular and subcellular membranes, has been explained through different perspectives, including electrostatic repulsion [25, 29–31], configurational barriers [7], desolvation penalties [32] and disruption of the highly constrained water structure at the SF [33]. However, the evidence suggests that several or all these factors occur concurrently in the process. At the molecular level, a reorientation of the water molecules inside the pore passage takes place thanks to the presence of two asparagine residues facing each other in the NPA motif, which create H-bonds with the water molecule in transit, substituting some of the water-water H-bonds [25, 30]. The hydrogen bonds with the two Asn residues, instead, facilitate water diffusion through the pore [18, 26]. This reorientation allows their continuous flux so that, when one molecule moves through the channel, the next one is readily available to follow the same path [13, 25, 26, 30]. On the other hand, water molecules in the SF region exhibit highly correlated motions among themselves—dictated by protein-water H-bond interactions—while their connectivity to water molecules outside the SF becomes weaker or more discontinuous *[33]*. Moreover, a slightly different mechanism has been suggested in some aquaporins. For instance, in hAQP4, the two asparagines (Asn97 and Asn213) separately form H-bonds with two different water molecules, while in most other AQ(G)Ps only one water molecule is involved at a time [20]. This arrangement has also been observed in the crystal structures of hAQP5 [19] and SoPIP2;1 [20]. In aquaporins 0, the Asn184 donate a hydrogen bond to a tyrosine, Tyr23, which in turn donates it to the water molecule [34, 35]. In this regard, the low water permeation observed in these AQPs (see Table S1) could be associated with the presence of this Tyr23, which is not present in other aquaporins.

In addition to the key role previously described, the NPA region can also function as a selectivity filter. When the two NPA motifs approach each other, a constriction forms, although the narrowest point most often occurs at the ar/R region.

Little sequence variation is found in the NPA motif through the whole family of aquaporins (Table S4). This occurs, for instance, in human aquaglyceroporin 7 (hAQP7), which allows glycerol permeation through cell membranes [36]. Human AQP7 exhibits uncommon substitutions in its NPA motifs while both asparagine residues are conserved. This results in the formation of distinctive triads: NAA (Asn94, Ala95, and Ala96) and NPS (Asn226, Pro227, and Ser228, see Table S4) [36]. Another case pertains to the aquaglyceroporin found in the malarial parasite *Plasmodium falciparum* (pAqgp), a protein of great significance in the life cycle of the parasite [37]. In pAqgp, both NPA regions undergo substitutions, although the two asparagines remain also conserved: NLA (Asn-Leu-Ala) and NPS (Asn-Pro-Ser) triads (Table S4). These modifications have been suggested to be crucial in maintaining the proper alignment of the asparagine residues at the selectivity filter [37].

*Selectivity Filter*. Aquaporins exhibit an hourglass structure, where the narrowest part of their pore is formed by some conserved amino acids—usually in the number of four—that facilitate the passage of water [26, 38] in single file [9]. The composition of this region, known as selectivity filter (SF) or aromatic/arginine (ar/R) motif [5, 39], affects the dimension and polarity of the pore, ultimately influencing its solute selectivity and permeation rate [13, 14, 23, 40]. SF contains at least one arginine and one aromatic residue, although the specific composition may vary among different AQ(G)Ps (see Table S4).

In orthodox aquaporins, there is also a highly conserved His residue, which normally faces the positively charged Arg residue. This Arg provides two hydrogen bond donor sites to which the His may bind, constricting thus the pore passage [41]. The fact that this His appears replaced in AQGPs by a small Gly, preventing the formation of the key H-bond interactions mentioned, suggests an important role for the former in the solute selectivity of these proteins [26]. Krenc and colleagues [41], working with the rat aquaporin 1, demonstrated the importance of this His through a mutational study in the position.

Greater sequence variability is evident in the SF motif compared to the NPA sequences among the AQ(G)Ps listed in Table S4. In bAQP1, for instance, the SF is formed by a hydrophobic corner (Phe58) and a hydrophilic face (His182, Arg197 and Cys191) on the opposite side [12]. The same residues are found in hAQP1 forming an almost identical SF: Phe56, His180, Arg195 and Cys189 [23]. This SF sequence is also carried by the aquaporin 1 of other organisms (e.g., *Anabas testudineus*: aAqp1), as well as by human aquaporins 2 and 5 (hAQP2, hAQP5). This set of AQPs is indicated with the symbol ⊖ in Table S4.

Bovine and ovine aquaporins 0 (bAQP0, oAQP0), as well as the human and rat aquaporins 4 (hAQP4, rAQP4, with symbol ▵ in Table S4), present the same residues quartet in their SF motif, namely Phe, His, Arg and Ala. Three of these residues (Phe, His and Arg) match those found in the SF of the previous set of aquaporins (⊖), while the Cys is substituted by an Ala residue. In aquaporins 0, this change enables the formation of a secondary constriction region, resulting in reduced water permeation [42, 43]. In aquaporins 4, the His is thought to play a determinant role in preventing glycerol passage [20, 44].

Another subgroup of AQPs sharing the same residues in their SF motif, namely Phe, His, Arg and Thr (with symbol ⊡ in Table S4), includes the spinach plasma membrane aquaporin (SoPIP2;1) [45], the *E. coli* and *Agrobacterium fabrum* aquaporins Z (eAqpZ [13], aAqpZ2 [46]), and the *Arabidopsis thaliana* aquaporin AtPIP2;4 [47]. In this group, the distinctive element is the Thr residue in the fourth amino acid position. In the bacterial (*E. coli*) eAqpZ, for instance, the Arg residue (Arg189) has been suggested to be crucial, as it shows two diverse conformations that result in significant differences in water permeability [21]. In one of these conformations, Arg189 forms a hydrogen bond with Thr183, which is believed to prevent the passage of water [21]. Thus, eAqpZ can exhibit structures that are either permeable or impermeable to water.

Other sequences in the SF region show a Gly residue in the second position, with a variety of other amino acids in the first and fourth positions. This is the case, for instance, of the human aquaporins 7 (hAQP7) [36] and 10 (hAQP10) [48], as well as the AQGP from the malarial parasite *Plasmodium falciparum* (pAqgp) [37] and the *E. coli* glycerol-conducting channel (eGlpF) [12].

Ile or Ser residues can also accompany Arg in the central positions of the SF region (see mAqpM and aAqpM in Table S4). Archaeal (*Methanothermobacter marburgensis*) aquaporin (mAqpM) is particularly interesting because it may exhibit multi-functional transport features. In this aquaporin, the His residue, highly conserved among water-selective AQPs, is replaced by an isoleucine residue at position 187 (Phe62, Ile187, Arg202 and Ser196). This substitution, which creates in mAqpM a wider and more hydrophobic SF than that observed in aquaporins 1, is believed to be an adaptation for the conductance of larger and less polar permeants than water, e.g., $H_2S$ and $CO_2$, in addition to the water transport [15]. At the same time, the presence of Ile187—which together with Phe62 forms the hydrophobic part of the mAqpM's constriction region—creates a pore narrower than that of the *E. coli* glycerol-conducting channel, eGlpF, suggesting that mAqpM may be not suited to conduct glycerol [15].

A particular conformation of the SF has been identified in the ammonia-permeable *Arabidopsis thaliana* TIP2;1 [49] (His63, Ile185, Arg200 and Gly194), which is also permeable to water [50]. Here, the conserved His in position 2 is replaced by a more hydrophobic Ile185 increasing pore diameter. The most interesting feature, however, is the behavior of Arg200, whose sidechain is pushed to the side of the pore by His131, which acts as a fifth element of the SF motif. Besides, Arg200 is H-bound to His63, conserved across all tonoplast intrinsic proteins (TIPs), which replaces the aromatic residue (Phe or Trp) mostly present in the other AQPs. The presence of both His63 and Arg200, along with the referred SF's fifth residue, His131, are a characteristic element of aquaammoniaporins [50].

In summary, three residues—Phe, His, and Arg—are conserved across the majority of AQ(G)Ps analyzed in this study. Notably, Arg is conserved across all AQ(G)Ps. While His is highly conserved among all orthodox AQPs [12, 16, 17, 20, 21, 46, 50], this hydrophilic residue is replaced by a Gly in AQGPs [9, 36, 37, 48], leading to increased hydrophobicity and pore size.

**Pairwise Comparison (Structure and Sequence)**

Systematic structural similarity (Template Modeling, TM-score) and sequence identity (SI) pairwise analyses have been carried out for the AQ(G)Ps studied in this work. TM-score is a length-independent metric for assessing the topological similarity of two protein structures. It ranges between 0 and 1, where 1 indicates a perfect match between two structures, values below 0.17 correspond to randomly chosen unrelated proteins whereas structures with a score higher than 0.7 assume generally the same fold in SCOP/CATH [51, 52]. Such length independency has prompted the recommendation of this metric over the more commonly used RMSD metric for global comparative structural analyses [51].

The pairwise structural comparison between different proteins was carried out using the Pairwise Structure Alignment tool, available at the RCSB PDB website [53]. The selected protein chains are superposed based on the position of the C-alpha atoms of the backbone and on the location of secondary structural elements. Here, we selected the initial chain of the tetrameric AQ(G)Ps' structures for the calculations. The tool provides several alignment methods, both rigid and flexible. Here we adopt a rigid body alignment, which is suitable for overall structural comparison of proteins which have closely related 3D shapes. Particularly, we adopt the FATCAT algorithm [54, 55], which allows for flexible structure comparison but can be modified to a rigid structure comparison considering only alignments with matching sequence order. The method allows to calculate the TM-score [52] and the percentages of sequence identity per protein pair, among other features.

Results of the pairwise analyses carried out are shown in Fig. S2. The pairwise TM-scores depicted on the 2D map (Fig. S2a: > 0.75 for all pairs) confirm the high degree of structural similarity among all the AQ(G)Ps, regardless the live kingdom or the class or sub-class they belong to. TM-score average per aquaporin (computed across all structures, as shown in Fig. S2b) were calculated. The averages consistently surpass the value of 0.8, with the exception of *Komagataella pastoris* aquaporin 1 (0.77), that is the only fungal AQP and the most structurally distinct protein within the analyzed AQ(G)Ps set. This distinctness is also evident from the pairwise TM-score-based clustering carried out and shown in Fig. S3.

Despite the high global structural similarity among all types of AQ(G)Ps (Fig. S2b), the specific structural features exhibited by the AQ(G)Ps allow them to be grouped by the kingdom of life (e.g., animals, plants, protista, fungi, monera) or by aquaporin type (AQPs or AQGPs; see Table 1 of the main text), as shown in Fig. S3. Indeed, AQPs pairs showing the highest pairwise TM-score (> 0.95) are the *Archaeoglobus fulgidus* probable aquaporin M (aAqpM) and the *Methanothermobacter marburgensis* aquaporin M (mAqpM), with a TM-score of 0.99 (71 % SI, Fig. S2c), the *Agrobacterium fabrum aquaporin* Z2 (aAqpZ2) and the *Escherichia coli* aquaporin Z (eAqpZ), with a TM-score of 0.97 (74 % SI), the hAQP1 and the bAQP1, with a TM-score of 0.96 (89 % SI), the bAQP0 and oAQP0, with a TM-score of 0.96 (98 % SI). Consistent with this, the human aquaglyceroporins hAQP7 and hAQP10 are grouped with the other two AQGPs in the set, namely pAqgp (*Plasmodium falciparum*) and eGlpF (*E. coli*), despite belonging to different kingdoms of life. All against all, mammalian AQPs shows the highest pairwise TM-score averages (Fig. S2b), namely oAQP0 (0.856), hAQP4 (0.854), rAQP4 (0.853), and bAQP0 (0.853).

At the sequence level, the highest pairwise SI averages (calculated over all the AQ(G)Ps) are exhibited by the orthodox AQPs from animals, namely hAQP1 (0.415), hAQP5 (0.411), oAQP0 (0.41), bAQP0 (0.41), hAQP4 (0.41), rAQP4 (0.41), bAQP1 (0.408), hAQP2 (0.405) and aAqp1 (0.405) (Fig. S2d).

**Pore Shape Analysis**

Pore diameter is perhaps the most comprehensive physical parameter to describe the size and shape of an aquaporin's pore. The PoreWalker tool [56] facilitates the calculation of pore diameter profiles and other properties, such as pore regularity. Additionally, it provides detailed representations of longitudinal cuts of the pores, along with the list of amino acids that make up the pore lining. Fig. 4a and left top panels in Fig. S1 depict the pore diameter profile for the 22 AQ(G)Ps analyzed in this work. Table S3 contains detailed information including the minimum diameter exhibited by the AQ(G)Ps' pore. Calculated pore diameters by PoreWalker tool [56] range from 1.4 to 26.5 Å along the entire longitudinal axis among all the AQ(G)Ps listed in Table 1 of the main text. Diameter profiles indicate that both the cytoplasmic and extracellular pore vestibules have, in general, a conical shape that extends to the single-file region (constriction site) located at the center of the membrane (Fig. 4b, Fig. 4f and Figure S1). This hourglass shape, however, varies from aquaporin to aquaporin, being highly regular in bovine and ovine aquaporins-0 (bAQP0 81% and oAQP0 75% regularity along the pore, Table S3), aquaporins-1 (66-85%), human aquaporin-2 (70%), and human aquaporin-4 (65%), whereas it is less regular in the rest of AQ(G)Ps, namely: aquaporins from plants (39-58%), bacteria (30-64%), and aquaglyceroporins (41-61%, Table S3). Occasionally, the loss of regularity in the pore come together with the loss of the conical shape in one of the two halves, leading to a funnel-shaped pore. This clearly occurs with the two plant AQPs from the highly conserved subfamily known as PIPs, SoPIP2;1 and AtPIP2;4, as well as the aquaglyceroporin hAQP10 (Table S3 and Fig. S1).

PIP aquaporins have been extensively studied and shown to exhibit well-conserved auto-regulatory molecular-level mechanisms known as 'pore gating' (closure mediated by conformational change of the cytoplasmic loop D, dependent of pH, post-translational modification or calcium-binding) [57], allowing them to adjust membrane water permeation under diverse conditions [45, 58–62]. Indeed, the selected crystal structure of SoPIP2;1 (PDB ID 1Z98 [45]) has been shown to be a closed (pore-gated) conformation [45], which is also the case of AtPIP2;4 (verified by structure comparative analysis with SoPIP2;1, see Fig. S6). Likewise, a pH-dependent pore gating mechanism has been identified in the human aquaglyceroporin hAQP10 [48], wherein His80 has been suggested to be the key residue (pH-sensor). The crystal structure of hAQP10 here analyzed (PDB ID 6F7H [48]) exhibits the closed (pore-gated) conformation (Fig. S6). Taking all this into account and examining the diameter profiles calculated by PoreWalker [56] for these AQPs, it seems that the apparent pore deformation from a classical hourglass (into a funnel-like shape), as predicted for SoPIP2;1, AtPIP2;4 and hAQP10, is mainly due to the pore closure at the cytosolic interface (Fig. S6). This regulatory mechanism seems to be less conserved in animal orthodox AQPs. Nonetheless, evidence suggesting the presence of other regulatory mechanisms, such as mechanosensitivity [63] and voltage-gating [64–66] have also been described. A recent literature review on the pore gating phenomena in AQPs is available in Ref. [57].

Water-selective AQPs typically exhibit a minimum pore diameter slightly smaller than that of a water molecule (~2.8 Å), while AQGPs have larger diameters [21] (Table S3). In most AQ(G)Ps, the constriction site (the minimum diameter) of the pore is located at or very close to the SF motif (see Table S3, Fig. 4a and Fig. S1). In contrast, *Anabas testudineus* aquaporin-1 (aAqp1) and *Arabidopsis thaliana* TIP2;1 aquaporin (AtTIP2;1) exhibit the minimum diameter near the NPA motifs. In hAQP10 and AtPIP2;4 the minimum diameter is located near the cytoplasmic border of the pore (Table S3), consistent with the occurrence of a pore closed (pore gating)—as described above (Fig. S6). Among the water-selective AQPs, the aquaporins-0 (bAQP0 and oAQP0) and the human aquaporin-5 (hAQP5) exhibit the smallest diameters, ranging from 1.4 to 1.7 Å. This correlates well, particularly for aquaporins-0, with their $p_f$ values, which are the lowest among all AQ(G)Ps (Table S1). However, as previously reported [67], and as further confirmed here by cross-referencing the data in Table S3 with those in Tables S1 and S2, the minimum pore diameter for the AQ(G)Ps analyzed does not follow the same hierarchical order as the experimental permeation rates. Specifically, the relative minimum pore diameter order is: oAQP0 < bAQP0 ~ hAQP5 < eAqpZ < kAqp1 ~ aAqpM ~ aAqpZ2 < AtPIP2;4 ≈ SoPIP2;1 ≈ hAQP2 < rAQP4 < bAQP1 < mAqpM ≈ AtTIP2;1 < hAQP4 ≈ aAqp1 < eGlpF < pAqgp < hAQP1 ≈ hAQP7 << OsNIP2;1 << hAQP10 (Table S3). In contrast, the relative $p_f$ order for the AQ(G)Ps that have been experimentally measured (average or individual values, as reported in Table S1 and showed in Fig. 2 of the main text) is: bAQP0 ≈ oAQP0 < mAqpM < eGlpF < bAQP1 < hAQP1 < eAqpZ < hAQP2 < hAQP4 << rAQP4. Likewise, for the AQ(G)Ps with corrected water permeabilities reported ($p_f^{corr}$, Table S2 and Fig. 2 of the main text) the relative order differs, namely: oAQP0 < sAqp < rAQP8 < hAQP2 < hAQP1 < eAqpZ ≈ rAQP4 < mAQP11 < hAQP4 < bAQP1 < RsAqpZ < eGlpF.

**Pore Composition Analysis**

Although specificities of the NPA and the SF motifs of AQ(G)Ps have been, in general, described, a comprehensive amino acid composition analysis of the AQ(G)Ps' pore lining at the widest available 3D structural level seems to be missing. Here, using the pore lining amino acid composition provided by PoreWalker [56] (**Supplementary Files**), we extracted the per amino acid and per amino acid type statistics (i.e., non-polar, polar uncharged, polar positively charged, and polar negatively charged residues in the pore lining) across all the AQ(G)Ps studied, and compared them with the amino acidic composition for the whole protein (Table S7 and Fig. S4).

Composition averages per amino acid type for all investigated proteins indicate that the percentages of non-polar (i.e. Gly, Ala, Val, Ile, Leu, Met, Pro, Phe, Trp) and negatively charged (Asp, Glu) amino acids are reduced by 3.4% (p-value=0.0019) and 1.4% (p-value=0.0008), respectively) in the pore lining compared to the whole protein (Fig. S4). Conversely, polar uncharged (Ser, Thr, Cys, Tyr, Asn, Gln) and positively charged (Arg, Lys, His) residues are prevalent by 2.2% (p-value=0.0068) and 2.6% (p-value=0.0003), respectively) in the pore lining. The majority of the AQ(G)Ps here analyzed follows this general pattern individually (see Table S7). Only hAQP7, one out of the four AQGPs in the group, presents a totally opposed profile, namely higher percentages of non-polar and negatively charged amino acids while lower percentages of polar uncharged and positively charged residues in the pore lining. hAQP10—another AQGP—, and OsNIP2;1—a silicon influx transporter in the *Oryza sativa japonica* plant—, do not either fully meet the general composition picture presented by the rest of AQ(G)Ps.

Focusing solely on the pore lining residues, orthodox AQPs from the animal kingdom generally exhibit a higher content of polar residues (charged and uncharged, >38-40%, Table S7) compared to AQPs from other kingdoms, and to hAQP7 (34.4%) and hAQP10 (30.3%), the only two mammalian AQGPs in the set. The amino acid contents of these two AQGPs more closely match those of non-animal AQPs (Table S7). The higher total content of polar residues in orthodox AQPs from animals relates to the higher content of polar uncharged residues (23.3-33.3% vs. 17.0-29.2% in the remaining AQ(G)Ps) and positively charged residues (10.0-18.0% vs. 4.8-12.9%). The content of negatively charged residues is similar in both subgroups (2.1-6.7% vs. 2.0-7.5%). All in all, the current composition analysis on the reduced sample of AQ(G)Ps here studied reveals clear differences in amino acid type composition within the pore lining of animal AQPs compared to AQGPs. Such differences are not as evident in the rest of AQPs or AQGPs from the other kingdoms. Additionally, the analysis shows that only the AQGPs (hAQP7, hAQP10, pAqgp and eGlpF) have a pore lining composition where the content of non-polar amino acids is similar to or higher than that in the whole protein (Table S7), which could be related with their different selectivity and permeability (Table S1) compared to orthodox AQPs.

# SUPPLEMENTARY TABLES

## Table S1. Experimental or computed water permeation rate(s) reported for AQ(G)Ps.

| AQ(G)P[a] (PDB ID) | Experimental Osmotic Permeability, $p_f$[b] (x $10^{-14}$ $cm^3 \cdot s^{-1}$) | Reference | Comments[c] | Computed Osmotic Permeability, $p_f$[b] (x $10^{-14}$ $cm^3 \cdot s^{-1}$) | Reference | Comments[c] |
|---|---|---|---|---|---|---|
| bAQP0 (1YMG) | <0.4 | [68] | PSA: proteoliposomes prepared by reverse-phase evaporation, by adding phosphatidylcholine (PC, 400 mg), phosphatidylinositol (PI, 40 mg), and cholesterol (240 mg), (moles of PC:PI:cholesterol = 11:1:11); Buffer: EDTA-phosphate (pH 7.4); Yield of bAqp0 in proteoliposomes: 33%; Temperature of measurement (T): not specified (NS). | 0.2±0.2 | [67] | Measurements from equilibrium MDS of the protein in a lipid bilayer (5 ns x 2 replicas, NPT); Starting structure PDB: 1YMG; FFs: Charmm22+CMAP (protein), Charmm27 (lipids); WM: Tip3p. |
| | 0.028 | [69] | OSA: measurements performed in *Xenopus laevis* oocytes injected with cRNAs coding for the protein; Buffer: 96 mM NaCl, 2 mM KCl, 1 mM $MgCl_2$, 1.8 mM $CaCl_2$, 5 mM HEPES (pH 7.6); T: NS. | | | |
| | 0.015 | [70] | OSA: measurements performed in Xenopus laevis oocytes injected with cRNAs coding for the protein; Buffer: 96 mM NaC1, 2 mM KC1, 1.8 mM $CaC1_2$, 1 mM $MgC1_2$ and 5 mM HEPES (pH 7.4) plus supplements (2.5 mM sodium pyruvate, 10 µg/ml aminobenzylpenicillin, 10 µg/ml streptomycin); T: 10$^{o}$C. | | | |
| | **Ave±SEM[*]: <0.15±0.13** | | | | | |
| oAQP0 (2B6O) | 0.2±0.05 | [71] | PSA: proteoliposomes reconstituted by dialysis by mixing purified oAqp1 (0.5 mg/mL) with the lipids PC, PS (phosphoserine) with cholesterol (1:4:5) | 0.2±0.2 | [71] | Measurements from equilibrium MDS of the protein in a POPC lipid bilayer (500 ns, NPT, 305 K, 1 atm); Starting structure PDB: 2B6O; FFs: Charmm36 (protein & |
| | | | solubilized in OG at a lipid-protein ratio of 2 mg/mg; Buffer: 20 mM HEPES, 100 mM NaCl (pH 7.5). Measurements performed under a pressure gradient created by osmotic agents; T: 15ºC. | | | lipids); WM: --; 150 mM NaCl; Essential-dynamics (ED) simulations were also carried out to restrain the protein in the open and closed states. |
| rAQP0 (no PDB structure) | 0.3±0.1 | [72] | OSA: measurements performed in *Xenopus laevis* oocytes injected with cRNAs coding for the protein; Buffer: Barth's solution (pH 7.4); T: 10ºC. | | | |
| bAQP1 (1J4N) | 2.3 | [73] | PSA: proteoliposomes prepared by reverse-phase evaporation mixing purified bAqp1 and the lipids PC, PI with cholesterol (11:1:11); Buffer: 1 mM EDTA, 100 mM Na-phosphate (pH 7.4) and 200 mM octyl glucoside (OG). Measurements performed under osmotic gradient created by different solute concentrations of the Buffer (from 10 to 60-100 mM Na-phosphate); T: NS. | 10.1±4.0 | [67] | Same MDS setup as for bAQP0; Starting structure PDB: 1J4N. |
| | | | | 7.1±0.9 | [74] | MDS of the protein in a lipid bilayer (5 ns x 4 replicas, NVT, 310 K); A water pressure gradient is settled by applying a constant force on all water molecules in the main direction of the channel; Starting structure PDB: 1J4N; FFs: Charmm22+CMAP (protein), Charmm27 (lipids); WM: Tip3p. |
| | | | | **Ave±SEM: 8.8±1.7** | | |
| hAQP1 (7UZE) | 6.0±1.0 | [68] | Same as for bAQP0. Yield of bAQP0 in proteoliposomes: 50%. | 7.5±0.5 | [25] | Measurements from equilibrium MDS of the protein in a lipid bilayer (10 ns, 2fs time-step, NPT, 300 K); Starting structure PDB: 1H6I; FFs: Gromos (protein & lipids). |
| | 1.2 | [69] | Same as for bAQP0. | | | |
| | 1.4 | [70] | Same as for bAQP0. | | | |
| | 3.4 | [73] | Same as for bAQP1. | | | |
| | 6.8 | [75] | PSA: measurements performed under osmotic gradient produced by manitol solutions; Buffer: Tris/HCl (pH 7.4); T: 10ºC. | | | |
| | 11.7±1.8 | [76] | PSA: measurements performed under osmotic gradient with sucrose solutions on proteoliposomes prepared by mixing purified hAqp1, *E. coli* phospholipid, OG, and Tris-HCl; Buffer: MOPS (pH 7.5); Yield of hAQP1 in proteoliposomes: 50%; T: NS. | | | |
| | 5.43 | [77] | Same as done by Zeidel, et al. [76] | | | |
| | 4.6 | [78] | Same as done by Zeidel, et al.[76], except that the concentration of hAQP1 in the proteoliposomes was varied in a range of values to determine the unit water conductance more precisely. | | | |
| | **Ave±SEM: 5.1±1.2** | | | | | |
| rAQP1 (no PDB structure) | 6.0 | [72] | Same as for rAQP0. | | | |
| rabAQP1 (no PDB structure) | 2.6 | [73] | Same as for bAQP1. | | | |
| pAQP1 (no PDB structure) | 1.9 | [73] | Same as for bAQP1. | | | |
| oAQP1 (no PDB structure) | 2.0 | [73] | Same as for bAQP1. | | | |
| hAQP2 (4NEF) | 9.3±0.3 | [79] | PSA: measurements performed under osmotic gradient with sucrose solutions on proteoliposomes prepared by mixing | 5.3±2.2 | [80] | MDS of the protein in a lipid bilayer (20 ns x 1 replica, NVT, 310 K); Starting structure PDB: 4NEF; FFs: |
| | | | purified hAQP2 and E. coli lipids; Buffer: 20 mM Tris-HCl, 50 mM NaCl (pH 7.2); T: NS. | | | Charmm22+CMAP (protein), Charmm27 (lipids); WM: Tip3p. |
| rAQP2 (no PDB structure) | 3.3±0.2 | [72] | Same as for rAQP0. | | | |
| hAQP4 (3GD8) | 11.0±1.0 | [81] | PSA: measurements performed under osmotic gradient with sucrose solutions on proteoliposomes prepared by mixing purified hAQP4 and a mixture of POPC, POPG and cholesterol (2:1:2); Buffer: 20 mM HEPES, 200 mM NaCl (pH 8.0); T: 4ºC. | 1.3±0.3 to 9.5±2.1 | [82] | Same approach followed by Hashido et al. [67] (200 ns x 1 replica, NPT, 310 K); Starting structure PDB: 3GD8. |
| | | | | 2.9±0.5 | [64] | Equilibrium MDS of the protein in a lipid bilayer (100 ns x 1 replica, NPT, 298 K); Starting structure PDB: 3GD8; FFs: Charmm22+CMAP (protein), Charmm27 (lipids); WM: Tip3p. |
| | | | | **Ave±SEM: 4.2±1.3** | | |
| rAQP4 (2D57) | 24.0±0.6 | [72] | Same as for rAQP0. | 7.4±2.6 | [83] | Same MD setup as for bAQP0 by Hashido et al[67]; Starting structure PDB: 2D57; Introduction of a $p_f$-matrix method for analysis, where $p_f$ is decomposed into contributions from each channel local region. |
| | 9.0 to 35.0 | [84] | PSA: measurements performed under osmotic gradient with sucrose solutions on proteoliposomes prepared by mixing purified rAQP4, octyl glucoside and four different mixtures of two phospholipids: 1) dierucoylphosphatidylcholine & dielaidoylphosphatidylglycerol, 2) dieicosenoylphosphatidylcholine & dielaidoylphosphatidylglycerol, 3) palmitoyloleoylphosphatidylcholine & dielaidoylphosphatidylglycerol, | 25.0±1.0 (1) 33.0±1.0 (2) | [84] | MDS of the protein in two lipid bilayer of different thickness: 1) dimyristoylphosphatidylcholine, 2) dierucoylphosphatidylcholine (20 replicas each, 40ns total, NVT, 300 K); r-RESPA multiple-time-step integrator applied with time steps of 2 and 4 fs for short-range non-bonded and long-range electrostatic interactions, respectively; Water pressure gradient settled by applying an ion gradient (50 mM NaCl on the cytoplasmic side and 150 mM on the |
| | | | 4) ditridecanoylphosphatidylcholine & dimyristoylphosphatidylglycerol; Buffer: 25 mM HEPES, 50 mM NaCl, pH 7.4; T: $20^{o}$C. | | | extracellular side); Starting structure PDB: 2D57; FFs: Charmm36 (protein & lipids); WM: modified Tip3p; 200 mM NaCl. |
| | **Ave±SEM: 23.0±1.0** | | | **Ave±SEM: 21.8±7.6** | | |
| hAQP5 (3D9S) | | | | 5.1 | [85] | Same approach followed by Hashido et al. [67] (500 ns x 1 replica, NPT, 300 K); Starting structure PDB: 3D9S. |
| rAQP5 (no PDB structure) | 5.0±0.4 | [72] | Same as for rAQP0. | | | |
| hAQP7 (6QZI) | | | | 4.8±0.2 | [36] | MDS of the protein in a lipid bilayer (500-1000 ns x 5 replicas, NPT, 300 K); Starting structure PDB: 6QZI; FF: Charmm36m for protein and lipids; WM: Tip3p. |
| mAqpM (2F2B) | 0.6 | [86] | PSA: measurements performed under osmotic gradient with sucrose solutions on proteoliposomes prepared by mixing purified mAqpM, octyl glucoside and *E. coli* phospholipids; Buffer: MOPS (pH 7.5); T: $4^{o}$C. | 6.4±1.4 | [40] | MDS of the protein in a lipid bilayer (20ns x 1 replica, NPT, 300 K); Starting structure PDB: 6QZI; FF: Charmm36m for protein and lipids; WM: Tip3p. |
| | 0.7 | [16] | PSA: measurements performed by setting an osmotic gradient with sucrose solutions on proteoliposomes; Buffer: 1.2% octyl glucoside, 50 mM HEPES, 100mM NaCl (pH 7.4); T: $12^{o}$C. | | | |
| | **Ave±SEM: 0.7±0.1** | | | | | |
| eAqpZ (1RC2) | >=10.0 | [87] | PSA: measurements performed under osmotic gradient with sucrose solutions on proteoliposomes prepared by mixing purified AqpZ, octyl glucoside and lipids; Buffer: MOPS (pH 7.5); T: NS. | 3.2±0.7 to 4.4±0.7 | [88] | MDS of the protein in several lipid bilayer compositions (18.5-26.5ns x 1 replica, NPT, 310 K); Starting structure PDB: 1RC2; FFs: Charmm22+CMAP (protein), Charmm27 (lipids); WM: Tip3p. |
| | 2.0 | [89] | PMA: measurements performed after forming planar lipid bilayers from unilamellar vesicles (proteoliposomes) containing purified AqpZ. Transmembrane water flux measured from solute concentration changes in the immediate membrane vicinity; 50 mM MOPS, 100 mM NaCl, 0.3, µM $CaCl_2$ (pH 7.5); T: NS. | 15.9±5.1 | [67] | Same as for bAQP0; Starting structure PDB: 1RC2. |
| | **Ave±SEM: >6.0±4.0** | | | **Ave±SEM: 8.2±3.9** | | |
| eGlpF (1FX8) | 2.0 | [90] | PSA: measurements performed under osmotic gradient with sucrose solutions on proteoliposomes prepared by mixing purified GlpF, octyl glucoside and lipids; Buffer: MOPS (pH 7.5); T: NS. | 14.0±0.4 | [91] | MDS of the protein in a lipid bilayer (4 replicas, 5ns total, NPT, 310 K); Pressure gradient settled by applying a constant force on all water molecules in the main direction of the channel; Starting structure PDB: 1FX8; FFs: Charmm22+CMAP (protein), Charmm27 (lipids); WM: Tip3p. |
| | 0.7 | [92] | Same as for AqpZ by Pohl et al. [89] | 16.0±3.0 | [67] | Same as for bAQP0; Starting structure PDB: 1LDI. |
| | | | | 8.6±1.1 to 13.1±3.4 | [88] | MDS of the protein in several lipid bilayer compositions (20-26ns x 1 replica, NPT, 310 K); Starting structure PDB: 1FX8; FFs: Charmm22+CMAP (protein), Charmm27 (lipids); WM: Tip3p. |
| | **Ave±SEM: 1.4±0.7** | | | **Ave±SEM: 13.6±1.5** | | |

[a] Experimental or computed water permeation rate(s) reported for the AQ(G)Ps listed in Table 1 of the main text, as well as for some AQPs without available 3D structure: rAQPx (rat aquaporins), rabAQP1 (rabbit aquaporin-1), pAQP1 (pig aquaporin-1) and oAQP1 (sheep aquaporin-1). [b] Experimental and computed (via MDS) permeabilities reported here refer to per-unit (pore channel) permeability. [c] Abbreviations: SEM, standard error of the mean; OSA, oocytes swelling assay; PSA, proteoliposome swelling assay; PMA, planar membrane assay; T, temperature of measurement; NS, not specified; MDS, molecular dynamics simulations; FF, force field; WM, water model; NPT, isothermal–isobaric ensemble; NVT, canonical ensemble.

**Table S2. Corrected single-channel water permeation rates for AQ(G)Ps.**[a]

| AQ(G)P (PDB ID) | Temperature[b] (°C) | Individual corrected single channel permeation rate[c], $p_f^{corr}$ (x $10^{-14}$ $cm^3 \cdot s^{-1}$) | Average of $p_f^{corr}$ (x $10^{-14}$ $cm^3 \cdot s^{-1}$) |
|---|---|---|---|
| bAQP1 (1J4N) | 5 | 32[d] | -- |
| eGlpF (1FX8) | 5 | 110[d] | -- |
| eAqpZ (1RC2) | 4 | 18 | 7.33[d] |
| | 5.5 | 5.86 | |
| | 6 to 8 | 4.20 | |
| | 5 | 1.24 | |
| hAQP1 (7UZE) | 10 | 6.80 | 4.6[d] |
| | 10 | 6 | |
| | 10 | 3.2 | |
| | 10 | 2.4 | |
| hAQP2 (4NEF) | RT | 3.27[d] | -- |
| hAQP4 (3GD8) | 4 | 11[d] | -- |
| rAQP4 (2D57) | 10 | 7.36[d] | -- |
| oAQP0 (2B6O) | 15 | 0.0874[d] | -- |
| rAQP8 | RT | 7.64[d] | -- |
| mAQP11 | NS | 1.2[d] | |
| RsAqpZ | NS | 44.4 | 44.7[d] |
| | NS | 45.0 | |
| sAqp | 20 | 0.32 | 0.35[d] |
| | 20 | 0.38 | |

[a] Corrected single-channel water permeation rates derived from experimental measurements at low-temperature (≤15°C, T in the second column), as reported by Wachlmayr *et al.* [93]. PDB entries (in parentheses) are provided for the eight AQ(G)Ps with experimentally resolved 3D structures, which were used as the QSAR training set. The remaining four AQ(G)Ps without resolved structures (no PDB entry): rAQP8 (rat aquaporin-8), mAQP11 (mouse aquaporin-11), RsAqpZ (*Rhodobacter sphaeroides*), sAqp (Soybean Aqp) served as an external test set for validation of Model 2.3b (see Table S6). [b] Temperature at which the water permeation rate was measured in the original functional assays, as reported in Table S1 of Wachlmayr *et al.* [93]. RT stands for room temperature, and NS stands for not specified. [c] Individual corrected water permeation corresponding to the reported measurement temperatures given in the second column. [d] Values used across the QSAR modeling (training or validation) described in the main text.

**Table S3. Pore characteristics and additional data extracted from PoreWalker output.**

| AQ(G)P[a] (PDB ID) | Min pore diameter[b] (Å) | Pore regularity[c] (%) | *z*-coord at min pore diameter[d] | Residues at min pore diameter[e] | Min pore diameter at SF[f] (Å) | *z*-coord at SF[g] | Min pore diameter at NPA[h] (Å) | *z*-coord at NPA[i] |
|---|---|---|---|---|---|---|---|---|
| bAQP0<br>(1YMG) | 1: 1.7<br>2: 1.7 | 80.5 | 1: -13.76<br>2: 4.24 | 1: I13, G60, I62, S63, G64, F75, F146, Y149, L157<br>2: V24, F48*, H172*, G182, R187*, A190, Y204 | 1.7 | F48: 3.46<br>H172: 6.11<br>A181: 6.62<br>R187: 2.46 | 2.7<br>2.5 | N68: -4.62<br>P69: -4.53<br>A70: -4.63<br>**N184: 0.00**<br>P185: -0.64<br>A186: - |
| oAQP0<br>(2B6O) | 1.4 | 75.0 | 3.22 | V24, G27, F48*, E134, G182, M183, R187*, Y204 | 1.4 | F48: 3.23<br>H172: 5.64<br>A181: 6.48<br>R187: 2.54 | 3.1<br>3.6 | N68: -4.71<br>P69: -<br>A70: -<br>**N184: 0.00**<br>P185: -<br>A186: - |
| bAQP1<br>(1J4N) | 2.5 | 69.0 | 2.13 | I25, I27, F58*, G59, A107, G181, I193, R197* | 2.5 | F58: 3.26<br>H182: 5.75<br>C191: 6.72<br>R197: 2.36 | 3.7<br>2.9 | N78: -4.36<br>P79: -<br>A80: -4.05<br>**N194: 0.00**<br>P195: -<br>A196: - |
| hAQP1<br>(7UZE) | 3.2 | 67.6 | -12.44 | G68, G72, L83, L154, T156, A168, V224, Y227 | 3.9 | F56: 3.24<br>H180: 4.99<br>C189: 6.16<br>R195: 2.55 | 3.8<br>3.9 | N76: -4.78<br>P77: -<br>A78: -4.66<br>N192: 0.00<br>P193: -<br>A194: - |
| aAQP1<br>(7W7S) | 2.7 | 85.0 | 0.15 | F23, A53, S99, V168, C169, L170, G171, I183, N184**, P185** | 3.9 | F50: 2.86<br>H172: 5.12<br>C181: -<br>R187: 2.28 | 3.5<br>2.7 | N70: -4.58<br>P71: -<br>A72: -3.86<br>**N184: 0.00**<br>P185: 0.07<br>A186: - |
| kAqp1<br>(3ZOJ) | 1: 1.9<br>2: 2.1 | 66.0 | 1: -17.25<br>2: 3.75 | Y27, P29, Y31, A124, I125, R129, V191, E192, Q261<br>L59, A62, F92*, H212*, I214, G222, R227* | 2.1 | F92: 3.74<br>H212: 5.45<br>A221: 6.62<br>R227: 2.93 | 3.7<br>2.7 | N112: -4.40<br>P113: -<br>A114: -<br>**N224: 0.00**<br>P225: -<br>A226: - |
| hAQP2<br>(4NEF) | 2.2 | 70.3 | 5.95 | G27, A45, A47, A101, C181*, S188 | 2.2 | F48: 3.72<br>H172: -<br>C181: 6.76<br>R187: 2.63 | 4.3<br>3.0 | N68: -4.60<br>P69: -<br>A70: -<br>**N184: 0.00**<br>P185: -<br>A186: - |
| hAQP4<br>(3GD8) | 2.7 | 64.9 | 3.69 | V49, S52, F77*, G129, S211, R216* | 2.7 | F77: 4.03<br>H201: -<br>A210: 6.57<br>R216: 2.47 | 3.5<br>5.2 | N97: -4.60<br>P98: -<br>A99: -<br>**N213: 0.00**<br>P214: -<br>A215: - |
| rAQP4<br>(2D57) | 2.3 | 57.9 | 4.83 | V49, S52, F77*, G200, H201*, S211 | 2.3 | F77: 4.22<br>H201: 6.61<br>A210: 6.97<br>R216: 2.72 | 5.0<br>3.8 | N97: -4.54<br>P98: -<br>A99: -<br>**N213: 0.00**<br>P214: -<br>A215: -1.48 |
| hAQP5<br>(3D9S) | 1.7 | 33.3 | 4.55 | V25, F27, A48, F49*, H173*,<br>S183, R188* | 1.7 | F49: 3.76<br>H173: 5.56<br>C182: 6.60<br>R188: 2.94 | 5.0<br><br><br>3.8 | N69: -4.47<br>P70: -<br>A71: -<br>**N185: 0.00**<br>P186: -<br>A187: -0.91 |
| hAQP7<br>(6QZI) | 3.2 | 41.0 | 3.46 | M48, F74*, V211, Y223*,<br>A224, R229*, D330 | 3.2 | F74: 3.97<br>G214: -<br>Y223: 4.88<br>R229: 2.61 | 4.8<br><br><br>5.6 | N94: -4.35<br>A95: -<br>A96: -<br>**N226: 0.00**<br>P227: -<br>S228: - |
| hAQP10<br>(6F7H) | 1: 1.9<br><br><br>2: 4.6 | 51.2 | 1: -14.15[Δ]<br><br><br>2: 3.85 | 1: Q23, N75, G78, M89, L95,<br>A176, I177, L178, G184,<br>V185, P186, G188, Y260<br>2: M35, L37, T38, G62*, S63,<br>A65, A111, A114, G202*,<br>I211*, P212, R217*, D218 | 4.6 | G62: 4.88<br>G202: 5.46<br>I211: 4.54<br>R217: 2.63 | 4.9<br><br><br>4.8 | N82: -4.44<br>P83: -<br>A84: -<br>**N214: 0.00**<br>P215: -<br>A216: - |
| SoPIP2;1<br>(1Z98) | 2.2 | 39.5 | 4.41 | L52, I54, T55, F81*, G133,<br>H210*, T219*, G220, R225,<br>G228 | 2.2 | F81: 4.27<br>H210: 5.57<br>T219: 5.79<br>R225: 2.78 | 3.6<br><br><br>2.6 | N101: -4.31<br>P102: -<br>A103: -<br>**N222: 0.00**<br>P223: -<br>A224: - |
| mAqpM<br>(2F2B) | 2.6 | 39.5 | 1.28 | L18, F62*, A65, A111, V183,<br>I186, L198, N199**, R202* | 2.6 | F62: 2.97<br>I187: 5.38<br>S196: 5.95<br>R202: 2.43 | 4.8<br><br><br>3.5 | N82: -4.52<br>P83: -4.16<br>A84: -<br>**N199: 0.00**<br>P200: -<br>A201: - |
| aAqpM<br>(3NE2) | 2.0 | 64.3 | 1.02 | F63*, A66, L199, N200**,<br>R203*, P229 | 2.0 | F63: 2.16<br>I188: 4.63<br>S197: 5.57<br>R203: 2.33 | 4.5<br><br><br>3.4 | N83: -4.53<br>P84: -3.99<br>A85: -<br>**N200: 0.00**<br>P201: -<br>A202: - |
| OsNIP2;1<br>(7CJS) | 4.0 | 40.5 | 2.86 | V62, T65, I86, G88*, C204,<br>T216, S217, R222* | 4.0 | G88: 2.79<br>S207: 5.49<br>G216: 6.42<br>R222: 2.09 | 4.6<br><br><br>5.2 | N108: -4.50<br>P109: -<br>A110: -<br>**N219: 0.00**<br>P220: -<br>A221: - |
| AtTIP2;1<br>(5I32) | 2.6 | 56.4 | -4.58 | A70, V82, N83**, A85**,<br>Q108, L153, I180 | 3.6 | H63: 4.20<br>I185: 6.47<br>G194: 6.41<br>R200: 2.59 | 2.6<br><br><br>3.9 | N83: -4.46<br>P84: -<br>A85: -4.29<br>**N197: 0.00**<br>P198: -<br>A199: - |
| AtPIP2;4<br>(6QIM) | 1: 2.1<br><br>2: 3.1 | 57.5 | 1: -16.06$^{\Delta}$<br><br>2: 1.94 | 1: L114, F115, A117, V120,<br>R124, F186, T189, V202<br>2: F51, G88, M90, A136,<br>V215, I227, N228, R231*, | 3.4 | F87: 4.07<br>H216: 5.24<br>T225: 5.91<br>R231: 3.45 | 3.6<br><br><br>4.1 | N107: -4.69<br>P108: -<br>A109: -4.82<br>**N228: 0.00**<br>P229: -0.85<br>A230: -0.76 |
| eAqpZ<br>(1RC2) | 1.8 | 30.0 | 3.25 | V16, G19, F43*, H174*, S184,<br>R189*, S190 | 2.0 | F43: 3.02<br>H174: 4.93<br>T183: 6.05<br>R189: 2.59 | 3.6<br><br><br>3.5 | N63: -4.58<br>P64: -<br>A65: -<br>**N186: 0.00**<br>P187: -<br>A188: - |
| aAqpZ2<br>(3LLQ) | 2.0 | 39.0 | 4.58 | A40, F43*, I135, E136, H172*,<br>T181*, S182, S188 | 2.0 | F43: 3.25<br>H172: 5.55<br>T181: 6.32<br>R187: 2.45 | 5.1<br><br><br>3.0 | N63: -4.39<br>P64: -4.42<br>A65: -<br>**N184: 0.00**<br>P185: -<br>A186: - |
| pAqgp<br>(3C02) | 3.1 | 61.0 | 4.55 | M24, L48, W50*, L178, F190*,<br>A191, D197, P223 | 3.1 | W50: 5.08<br>G181: -<br>F190: 5.45<br>R196: 2.28 | 4.9<br><br><br>4.0 | N70: -4.34<br>L71: -<br>A72: -<br>**N193: 0.00**<br>P194: -<br>S195: - |
| eGlpF<br>(1FX8) | 3.0 | 48.6 | 3.59 | S45, W48*, A97, F200*, A201,<br>R206*, P236 | 3.0 | W48: 4.37<br>G191: -<br>F200: 4.13<br>R206: 2.78 | 4.9<br><br><br>6.1 | N68: -4.26<br>P69: -<br>A70: -<br>**N203: 0.00**<br>P204: -<br>A205: - |

[a] Aquaporin full names and their source organisms are given in Table 1 of the main text. [b] Minimum pore diameter calculated by the PoreWalker [56] along the *z*-coordinates (*z*-axis) of the AQ(G)Ps' pore. The *z*-axis refers to that perpendicular to the *xy* cellular membrane plane, i.e. the main/longest pore axis. In cases where a minimum pore diameter appears near the terminal ends of the pore (see diameter profiles in Fig. S1 and Fig. 4a of the main text), a second minimum closer to the central part is also reported. [c] Measure of regularity of the pore as issued by PoreWalker. The value refers to the percentage of pore centers (at 1Å step) that could be fitted to a line. [d] *z*-coordinates where the pore exhibits the minimum diameter(s). Values extracted from the diameter plots included in Fig. S1 and Fig.4a (left top panels), which were obtained at 3 Å step by PoreWalker. The *z*-coordinate associated to each amino acid is that of their beta carbon (βC). The *z*-coordinates along the whole *z*-axis were shifted to place the beta carbon (βC) of the asparagine (N) in the second NPA motif in the center of the *z*-coordinate system (*z*=0), for all the AQPs analyzed (check it out in the last column, in bold). This way the *z*-coordinates are aligned according to the equivalent residues in the AQPs (as if they were aligned structurally). With a delta (Δ) symbol appear those *z*-coordinates of minima located near the cytosolic or the extracellular interfaces of the pore. [e] Residues located within a *z*-coordinate range of ±2 Å of the pore minimum. With an asterisk (*) those amino acids forming the SF motifs. [f] Minimum pore diameter in the *z*-coordinates range encompassing the four residues forming the selectivity filter (SF). [g] *z*-coordinates of the amino acids (βC) forming the SF motif. [h] Minimum pore

diameter found within the *z*-coordinates range of the six residues forming the two NPA motifs. [i] *z*-coordinates of the amino acids (βC) forming the NPA motifs. In bold the asparagine in the second NPA motif that constitute the reference (*z*=0) coordinate. **Note:** All the information used to extract the data showed in this table is contained in the PoreWalker output files, provided as **Supplementary Files**. Pore diameter profiles and additional information on the pore shape are given more visually in Fig. S1 and Fig. 4a of the main text.

**Table S4. Sequences of the NPA and the selectivity filter (SF) motifs of selected AQ(G)Ps. The symbols highlight the difference or identity across the sequences of the NPA and the SF motifs.**

| AQ(G)P[a] | PDB ID | Sequences of the NPA motifs[b] | | | | Sequence of the SF motif[c] | | | | |
|---|---|---|---|---|---|---|---|---|---|---|
| bAQP0 | 1YMG | NPA 68-70 | & | NPA 184-186 | ⊖ | Phe48 | His172 | Arg187 | Ala181 | △ |
| oAQP0 | 2B6O | NPA 68-70 | & | NPA 184-186 | ⊖ | Phe48 | His172 | Arg187 | Ala181 | △ |
| bAQP1 | 1J4N | NPA 78-80 | & | NPA 194-196 | ⊖ | Phe58 | His182 | Arg197 | Cys191 | ⊖ |
| hAQP1 | 7UZE | NPA 76-78 | & | NPA 192-194 | ⊖ | Phe56 | His180 | Arg195 | Cys189 | ⊖ |
| aAQP1 | 7W7S | NPA 70-72 | & | NPA 184-186 | ⊖ | Phe50 | His172 | Arg187 | Cys181 | ⊖ |
| kAqp1 | 3ZOJ | NPA 112-114 | & | NPA 224-226 | ⊖ | Phe92 | His212 | Arg227 | Ala221 | △ |
| hAQP2 | 4NEF | NPA 68-70 | & | NPA 184-186 | ⊖ | Phe48 | His172 | Arg187 | Cys181 | ⊖ |
| hAQP4 | 3GD8 | NPA 97-99 | & | NPA 213-215 | ⊖ | Phe77 | His201 | Arg216 | Ala210 | △ |
| rAQP4 | 2D57 | NPA 97-99 | & | NPA 213-215 | ⊖ | Phe77 | His201 | Arg216 | Ala210 | △ |
| hAQP5 | 3D9S | NPA 69-71 | & | NPA 185-187 | ⊖ | Phe49 | His173 | Arg188 | Cys182 | ⊖ |
| hAQP7 | 6QZI | NAA 94-96 | & | NPS 226-228 | ⊡ | Phe74 | Gly214 | Arg229 | Tyr223 | ⊟ |
| hAQP10 | 6F7H | NPA 82-84 | & | NPA 214-216 | ⊖ | Gly62 | Gly202 | Arg217 | Ile211 | ⊙ |
| SoPIP2;1 | 1Z98 | NPA 101-103 | & | NPA 222-224 | ⊖ | Phe81 | His210 | Arg225 | Thr219 | ⊡ |
| mAqpM | 2F2B | NPA 82-84 | & | NPA 199-201 | ⊖ | Phe62 | Ile187 | Arg202 | Ser196 | ⊗ |
| aAqpM | 3NE2 | NPA 83-85 | & | NPA 200-202 | ⊖ | Phe63 | Ile188 | Arg203 | Ser197 | ⊗ |
| OsNIP2;1 | 7CJS | NPA 108-110 | & | NPA 219-221 | ⊖ | Gly88 | Ser207 | Arg222 | Gly216 | ⊕ |
| AtTIP2;1 | 5I32 | NPA 83-85 | & | NPA 197-199 | ⊖ | His63 | Ile185 | Arg200 | Gly194 | ⊠ |
| AtPIP2;4 | 6QIM | NPA 107-109 | & | NPA 228-230 | ⊖ | Phe87 | His216 | Arg231 | Thr225 | ⊡ |
| eAqpZ | 1RC2 | NPA 63-65 | & | NPA 186-188 | ⊖ | Phe43 | His174 | Arg189 | Thr183 | ⊡ |
| aAqpZ2 | 3LLQ | NPA 63-65 | & | NPA 184-186 | ⊖ | Phe43 | His172 | Arg187 | Thr181 | ⊡ |
| pAqgp | 3C02 | NLA 70-72 | & | NPS 193-195 | ▵ | Trp50 | Gly181 | Arg196 | Phe190 | ⊞ |
| eGlpF | 1FX8 | NPA 68-70 | & | NPA 203-205 | ⊖ | Trp48 | Gly191 | Arg206 | Phe200 | ⊞ |

[a] Full names and AQ(G)Ps' source organisms are given in Table 1 of the main text. [b] Information retrieved from the UniProtKB database [94]. [c] Information retrieved in some cases from the reviewed literature and in the others from structural alignment.

**Table S5. Resolution, experimental conditions and other relevant information on crystallization of AQ(G)Ps in the modeled datasets.**[a]

---

---

[a] The table is provided as an Excel **Supplementary File**. Data reported in this table has been retrieved from the Protein Data Bank [95] website ( 'Experiment' tab in each PDB entry).

**Table S6. Computed values of the PDs retained in Model 3.2b, along with predicted $p_f^{corr}$ values and residuals for the four AQPs used for external validation.**

| | | rAQP8 | mAQP11 | RsAqpZ | sAqp |
|---|---|---|---|---|---|
| PDs retained in Model 2.3b | $Regularity$[a] | 7.317 | 7.838 | 5.385 | 6.578 |
| | $N_H$[a] | 10 | 9 | 9 | 11 |
| | $Sites^{+}$[a] | 1 | 14 | 6 | 7 |
| Values used for validation of Model 2.3b | Predicted $p_f^{corr}$[b] | 2.05 | 14.98 | 35.93 | 2.28 |
| | (Pred. $ln\ p_f^{corr}$) | (0.72) | (2.71) | (3.58) | (0.83) |
| | Observed $p_f^{corr}$[b] | 1.2 | 7.64 | 44.7 | 0.35 |
| | (Obs. $ln\ p_f^{corr}$) | (0.18) | (2.03) | (3.80) | (−1.05) |
| | Pred. – Obs. non-log[b] | 0.85 | 7.34 | 8.77 | 1.93 |

[a] PD values were computed for the external AQ(G)Ps using the same procedures and tools described in the Materials and Methods for the training set, and were standardized using the training-set mean and standard deviation. [b] Values reported in the original scale of permeation rates (x $10^{-14}$ $cm^3 \cdot s^{-1}$).

**Table S7. Composition (in %) per amino acid (AA) and per amino acid type for the AQ(G)Ps' pore lining compared to whole protein.**

| AQ(G)P[a] (PDB ID) | Composition on…[b] | Non-polar AAs | | | | | | | | | | Polar AAs w/o charges | | | | | | | Polar AAs with (−) charges | | | Polar AAs with (+) charge | | | |
|---|---|---|---|---|---|---|---|---|---|---|---|---|---|---|---|---|---|---|---|---|---|---|---|---|---|
| | | Gly | Ala | Val | Ile | Leu | Met | Pro | Phe | Trp | Total | Ser | Thr | Cys | Tyr | Asn | Gln | Total | Asp | Glu | Total | Arg | Lys | His | Total |
| bAQP0 (1YMG) | Whole protein | 9.9 | 12.2 | 9.5 | 3.4 | 14.4 | 1.9 | 4.9 | 6.8 | 1.9 | 65.0 | 7.2 | 4.9 | 1.1 | 3.0 | 3.0 | 3.0 | 22.4 | 0.8 | 3.4 | 4.2 | 4.9 | 1.1 | 2.3 | 8.4 |
| | Pore lining | 11.1 | 11.1 | 8.9 | 4.4 | 12.2 | 3.3 | 3.3 | 5.6 | 1.1 | 61.1 | 7.8 | 4.4 | 0.0 | 4.4 | 5.6 | 1.1 | 23.3 | 1.1 | 2.2 | 3.3 | 4.4 | 2.2 | 5.6 | 12.2 |
| oAQP0 (2B6O) | Whole protein | 9.9 | 12.2 | 9.9 | 3.4 | 14.4 | 1.5 | 4.9 | 7.2 | 1.9 | 65.4 | 6.5 | 5.7 | 0.8 | 3.0 | 3.0 | 3.0 | 22.1 | 0.8 | 3.4 | 4.2 | 4.9 | 1.1 | 2.3 | 8.4 |
| | Pore lining | 8.8 | 9.9 | 9.9 | 3.3 | 12.1 | 3.3 | 3.3 | 5.5 | 0.0 | 56.0 | 5.5 | 4.4 | 1.1 | 5.5 | 6.6 | 3.3 | 26.4 | 0.0 | 2.2 | 2.2 | 9.9 | 0.0 | 5.5 | 15.4 |
| bAQP1 (1J4N) | Whole protein | 9.6 | 10.3 | 7.7 | 9.6 | 12.2 | 1.5 | 3.3 | 4.8 | 1.5 | 60.5 | 8.9 | 4.8 | 1.5 | 1.8 | 3.3 | 3.3 | 23.6 | 4.8 | 2.2 | 7.0 | 3.7 | 3.0 | 2.2 | 8.9 |
| | Pore lining | 13.3 | 10.0 | 7.8 | 10.0 | 11.1 | 1.1 | 3.3 | 4.4 | 0.0 | 61.1 | 5.6 | 5.6 | 2.2 | 1.1 | 6.7 | 5.6 | 26.7 | 2.2 | 0.0 | 2.2 | 5.6 | 0.0 | 4.4 | 10.0 |
| hAQP1 (7UZE) | Whole protein | 10.0 | 10.4 | 8.2 | 8.6 | 12.3 | 1.1 | 2.6 | 4.8 | 1.5 | 59.5 | 8.6 | 5.6 | 1.5 | 1.9 | 4.1 | 3.0 | 24.5 | 4.8 | 2.2 | 7.1 | 4.1 | 3.0 | 1.9 | 8.9 |
| | Pore lining | 12.4 | 9.0 | 7.9 | 6.7 | 10.1 | 0.0 | 1.1 | 4.5 | 0.0 | 51.7 | 5.6 | 6.7 | 2.2 | 2.2 | 4.5 | 4.5 | 25.8 | 3.4 | 1.1 | 4.5 | 11.2 | 3.4 | 3.4 | 18.0 |
| aAQP1 (7W7S) | Whole protein | 10.3 | 11.1 | 9.2 | 5.7 | 12.3 | 2.7 | 4.2 | 4.2 | 1.1 | 60.9 | 7.3 | 5.4 | 1.9 | 2.3 | 5.7 | 2.3 | 24.9 | 3.4 | 2.7 | 6.1 | 2.7 | 3.8 | 1.5 | 8.0 |
| | Pore lining | 8.6 | 9.7 | 7.5 | 7.5 | 9.7 | 2.2 | 3.2 | 4.3 | 0.0 | 52.7 | 11.8 | 3.2 | 3.2 | 2.2 | 8.6 | 4.3 | 33.3 | 1.1 | 2.2 | 3.2 | 3.2 | 4.3 | 3.2 | 10.8 |
| kAqp1 (3ZOJ) | Whole protein | 10.4 | 15.8 | 5.0 | 7.9 | 8.2 | 3.2 | 6.1 | 7.2 | 2.9 | 66.7 | 4.7 | 5.0 | 0.7 | 3.6 | 4.3 | 3.2 | 21.5 | 2.5 | 2.9 | 5.4 | 3.6 | 1.4 | 1.4 | 6.5 |
| | Pore lining | 12.9 | 17.2 | 6.5 | 6.5 | 8.6 | 0.0 | 3.2 | 6.5 | 3.2 | 64.5 | 2.2 | 5.4 | 1.1 | 4.3 | 7.5 | 4.3 | 24.7 | 1.1 | 3.2 | 4.3 | 3.2 | 2.2 | 1.1 | 6.5 |
| hAQP2 (4NEF) | Whole protein | 8.9 | 12.9 | 8.9 | 4.4 | 14.8 | 1.1 | 5.2 | 5.2 | 1.8 | 63.1 | 7.0 | 5.2 | 1.5 | 1.5 | 2.6 | 3.0 | 20.7 | 2.6 | 4.4 | 7.0 | 4.8 | 1.5 | 3.0 | 9.2 |
| | Pore lining | 9.1 | 14.1 | 8.1 | 7.1 | 7.1 | 1.0 | 3.0 | 6.1 | 1.0 | 56.6 | 9.1 | 4.0 | 3.0 | 1.0 | 5.1 | 2.0 | 24.2 | 3.0 | 3.0 | 6.1 | 6.1 | 2.0 | 5.1 | 13.1 |

| | | | | | | | | | | | | | | | | | | | | | | | | |
|---|---|---|---|---|---|---|---|---|---|---|---|---|---|---|---|---|---|---|---|---|---|---|---|---|
| hAQP4 (3GD8) | Whole protein | 10.5 | 8.7 | 10.2 | 8.0 | 8.0 | 3.1 | 3.7 | 5.3 | 2.2 | 59.8 | 6.5 | 5.9 | 2.5 | 2.2 | 2.8 | 2.5 | 22.3 | 3.4 | 4.3 | 7.7 | 3.4 | 4.6 | 2.2 | 10.2 |
| | Pore lining | 11.4 | 6.8 | 11.4 | 10.2 | 3.4 | 8.0 | 3.4 | 4.5 | 0.0 | 59.1 | 9.1 | 5.7 | 2.3 | 1.1 | 6.8 | 2.3 | 27.3 | 2.3 | 1.1 | 3.4 | 2.3 | 4.5 | 3.4 | 10.2 |
| rAQP4 (2D57) | Whole protein | 10.5 | 9.3 | 10.5 | 8.0 | 8.0 | 2.8 | 3.7 | 4.6 | 2.2 | 59.8 | 7.7 | 5.6 | 2.5 | 2.2 | 2.8 | 2.2 | 22.9 | 3.4 | 4.3 | 7.7 | 3.1 | 4.3 | 2.2 | 9.6 |
| | Pore lining | 12.2 | 6.7 | 11.1 | 10.0 | 3.3 | 6.7 | 0.0 | 5.6 | 1.1 | 56.7 | 12.2 | 5.6 | 2.2 | 1.1 | 4.4 | 1.1 | 26.7 | 3.3 | 3.3 | 6.7 | 2.2 | 3.3 | 4.4 | 10.0 |
| hAQP5 (3D9S) | Whole protein | 8.7 | 12.8 | 8.3 | 6.4 | 13.6 | 1.9 | 4.9 | 6.4 | 1.5 | 64.5 | 6.4 | 5.7 | 1.1 | 2.3 | 4.2 | 3.0 | 22.6 | 1.1 | 4.2 | 5.3 | 3.8 | 2.6 | 1.1 | 7.5 |
| | Pore lining | 8.5 | 9.6 | 6.4 | 10.6 | 9.6 | 1.1 | 4.3 | 8.5 | 0.0 | 58.5 | 7.4 | 5.3 | 2.1 | 3.2 | 7.4 | 3.2 | 28.7 | 1.1 | 1.1 | 2.1 | 6.4 | 2.1 | 2.1 | 10.6 |
| hAQP7 (6QZI) | Whole protein | 10.2 | 8.5 | 9.4 | 6.1 | 10.2 | 4.4 | 6.1 | 5.6 | 2.3 | 62.9 | 7.0 | 5.6 | 0.6 | 3.2 | 3.5 | 2.3 | 22.2 | 1.5 | 3.8 | 5.3 | 3.8 | 2.6 | 3.2 | 9.6 |
| | Pore lining | 9.7 | 8.6 | 12.9 | 8.6 | 7.5 | 5.4 | 5.4 | 6.5 | 1.1 | 65.6 | 4.3 | 5.4 | 1.1 | 5.4 | 4.3 | 1.1 | 21.5 | 2.2 | 3.2 | 5.4 | 4.3 | 1.1 | 2.2 | 7.5 |
| hAQP10 (6F7H) | Whole protein | 10.6 | 12.0 | 9.3 | 4.0 | 14.6 | 3.0 | 6.3 | 4.0 | 1.7 | 65.4 | 4.7 | 6.0 | 1.7 | 3.0 | 3.7 | 3.7 | 22.6 | 1.7 | 3.7 | 5.3 | 2.7 | 2.0 | 2.0 | 6.6 |
| | Pore lining | 11.1 | 13.1 | 11.1 | 7.1 | 11.1 | 4.0 | 7.1 | 4.0 | 1.0 | 69.7 | 4.0 | 5.1 | 1.0 | 4.0 | 3.0 | 3.0 | 20.2 | 2.0 | 3.0 | 5.1 | 3.0 | 0.0 | 2.0 | 5.1 |
| SoPIP2;1 (1Z98) | Whole protein | 10.7 | 13.5 | 9.3 | 7.5 | 8.5 | 1.8 | 5.0 | 7.1 | 2.1 | 65.5 | 5.3 | 5.0 | 1.4 | 3.2 | 3.2 | 2.1 | 20.3 | 2.5 | 2.5 | 5.0 | 2.8 | 3.9 | 2.5 | 9.3 |
| | Pore lining | 12.0 | 11.0 | 10.0 | 11.0 | 6.0 | 0.0 | 4.0 | 8.0 | 0.0 | 62.0 | 5.0 | 7.0 | 1.0 | 4.0 | 5.0 | 2.0 | 24.0 | 1.0 | 1.0 | 2.0 | 4.0 | 3.0 | 5.0 | 12.0 |
| mAqpM (2F2B) | Whole protein | 14.6 | 12.6 | 6.5 | 12.2 | 9.8 | 2.4 | 4.9 | 6.5 | 1.6 | 71.1 | 5.3 | 6.5 | 1.2 | 3.7 | 3.7 | 1.6 | 22.0 | 1.2 | 2.0 | 3.3 | 1.6 | 1.6 | 0.4 | 3.7 |
| | Pore lining | 16.2 | 12.1 | 6.1 | 15.2 | 8.1 | 2.0 | 4.0 | 3.0 | 1.0 | 67.7 | 4.0 | 9.1 | 1.0 | 3.0 | 3.0 | 1.0 | 21.2 | 2.0 | 2.0 | 4.0 | 3.0 | 3.0 | 1.0 | 7.1 |
| aAqpM (3NE2) | Whole protein | 15.0 | 14.6 | 7.7 | 11.0 | 10.2 | 2.8 | 5.7 | 6.1 | 1.6 | 74.8 | 2.8 | 6.1 | 0.4 | 3.3 | 3.3 | 1.2 | 17.1 | 1.6 | 2.4 | 4.1 | 2.4 | 1.2 | 0.4 | 4.1 |
| | Pore lining | 13.3 | 13.3 | 7.6 | 13.3 | 8.6 | 2.9 | 7.6 | 3.8 | 1.0 | 71.4 | 2.9 | 8.6 | 0.0 | 1.9 | 4.8 | 1.0 | 19.0 | 1.9 | 2.9 | 4.8 | 3.8 | 0.0 | 1.0 | 4.8 |
| OsNIP2;1 (7CJS) | Whole protein | 8.4 | 10.7 | 9.7 | 6.7 | 7.4 | 3.7 | 4.4 | 6.0 | 1.7 | 58.7 | 8.7 | 7.4 | 1.0 | 2.3 | 3.7 | 3.0 | 26.2 | 3.4 | 3.0 | 6.4 | 3.4 | 2.7 | 2.7 | 8.7 |
| | Pore lining | 15.7 | 14.3 | 11.4 | 5.7 | 4.3 | 7.1 | 2.9 | 2.9 | 0.0 | 64.3 | 5.7 | 8.6 | 1.4 | 1.4 | 2.9 | 2.9 | 22.9 | 2.9 | 1.4 | 4.3 | 4.3 | 0.0 | 4.3 | 8.6 |
| AtTIP2;1 (5I32) | Whole protein | 14.4 | 16.4 | 10.0 | 7.2 | 10.0 | 1.6 | 4.0 | 6.8 | 1.2 | 71.6 | 7.6 | 5.2 | 0.8 | 3.2 | 2.0 | 0.8 | 19.6 | 2.8 | 1.6 | 4.4 | 0.8 | 1.6 | 2.0 | 4.4 |
| | Pore lining | 8.3 | 16.7 | 9.7 | 6.9 | 8.3 | 2.8 | 4.2 | 1.4 | 0.0 | 58.3 | 11.1 | 8.3 | 0.0 | 4.2 | 4.2 | 1.4 | 29.2 | 2.8 | 1.4 | 4.2 | 2.8 | 1.4 | 4.2 | 8.3 |
| AtPIP2;4 (6QIM) | Whole protein | 11.7 | 13.7 | 8.6 | 7.2 | 7.6 | 1.7 | 5.2 | 7.2 | 1.7 | 64.6 | 4.5 | 5.2 | 1.7 | 4.5 | 2.7 | 1.7 | 20.3 | 4.1 | 2.4 | 6.5 | 3.8 | 3.4 | 1.4 | 8.6 |
| | Pore lining | 10.5 | 11.6 | 11.6 | 9.5 | 7.4 | 1.1 | 5.3 | 5.3 | 0.0 | 62.1 | 2.1 | 8.4 | 2.1 | 4.2 | 5.3 | 3.2 | 25.3 | 2.1 | 1.1 | 3.2 | 4.2 | 3.2 | 2.1 | 9.5 |
| eAqpZ (1RC2) | Whole protein | 15.2 | 14.7 | 9.5 | 7.4 | 11.3 | 1.3 | 3.9 | 7.8 | 2.2 | 73.2 | 5.2 | 4.8 | 0.9 | 2.2 | 1.7 | 1.3 | 16.0 | 1.3 | 3.0 | 4.3 | 2.2 | 2.2 | 2.2 | 6.5 |
| | Pore lining | 14.1 | 10.9 | 8.7 | 5.4 | 7.6 | 1.1 | 4.3 | 12.0 | 2.2 | 66.3 | 7.6 | 3.3 | 1.1 | 1.1 | 4.3 | 0.0 | 17.4 | 1.1 | 3.3 | 4.3 | 2.2 | 4.3 | 5.4 | 12.0 |
| aAqpZ2 (3LLQ) | Whole protein | 15.4 | 15.4 | 9.2 | 7.9 | 11.8 | 0.9 | 4.4 | 7.5 | 2.2 | 74.6 | 5.7 | 5.3 | 0.4 | 2.2 | 1.8 | 1.8 | 17.1 | 0.9 | 2.2 | 3.1 | 1.8 | 1.8 | 1.8 | 5.3 |
| | Pore lining | 13.6 | 12.5 | 10.2 | 8.0 | 8.0 | 0.0 | 3.4 | 11.4 | 1.1 | 68.2 | 4.5 | 4.5 | 1.1 | 1.1 | 4.5 | 1.1 | 17.0 | 1.1 | 2.3 | 3.4 | 3.4 | 3.4 | 4.5 | 11.4 |
| pAqgp (3C02) | Whole protein | 10.5 | 5.4 | 8.5 | 7.4 | 14.0 | 1.2 | 2.7 | 9.7 | 1.6 | 60.9 | 7.4 | 5.0 | 2.3 | 3.5 | 4.7 | 1.2 | 24.0 | 3.9 | 2.7 | 6.6 | 1.6 | 5.0 | 1.9 | 8.5 |
| | Pore lining | 8.6 | 7.5 | 8.6 | 8.6 | 12.9 | 1.1 | 3.2 | 5.4 | 3.2 | 59.1 | 6.5 | 5.4 | 1.1 | 2.2 | 4.3 | 1.1 | 20.4 | 4.3 | 3.2 | 7.5 | 3.2 | 6.5 | 3.2 | 12.9 |
| eGlpF (1FX8) | Whole protein | 11.4 | 12.1 | 9.3 | 7.5 | 10.7 | 2.1 | 5.3 | 7.5 | 1.8 | 67.6 | 4.3 | 5.3 | 2.1 | 2.5 | 2.5 | 2.5 | 19.2 | 2.8 | 3.2 | 6.0 | 2.5 | 2.8 | 1.8 | 7.1 |
| | Pore lining | 13.1 | 11.1 | 7.1 | 9.1 | 10.1 | 2.0 | 6.1 | 7.1 | 2.0 | 67.7 | 2.0 | 6.1 | 2.0 | 3.0 | 3.0 | 2.0 | 18.2 | 5.1 | 0.0 | 5.1 | 5.1 | 2.0 | 2.0 | 9.1 |
| **ALL PROTEINS** | **Whole protein** | **11.2** | **12.1** | **8.8** | **7.2** | **11.0** | **2.2** | **4.6** | **6.3** | **1.8** | **65.3±1.1** | **6.3** | **5.5** | **1.4** | **2.8** | **3.3** | **2.4** | **21.6±0.6** | **2.6** | **3.0** | **5.6±0.3** | **3.1** | **2.6** | **1.9** | **7.6±0.4** |
| | **Pore lining** | **11.6** | **11.2** | **9.1** | **8.4** | **8.5** | **2.6** | **3.9** | **5.8** | **0.9** | **61.8±1.2** | **6.2** | **5.9** | **1.5** | **2.8** | **5.1** | **2.3** | **23.8±0.9** | **2.1** | **2.0** | **4.2±0.3** | **4.5** | **2.4** | **3.4** | **10.2±0.7** |

[a] Aquaporins full names and their source organisms are given in Table 1 of the main manuscript. [b] Statistics for the whole protein composition obtained from the canonical aquaporin sequences retrieved from UniProtKB database [94], and statistics for the pore lining composition calculated based on the information given by PoreWalker [56]. FASTA files with the protein sequence from UniProtKB are provided as **Supplementary Files** along with all the output files issued by PoreWalker including the pore-lining amino acidic composition.

## SUPPLEMENTARY FIGURES

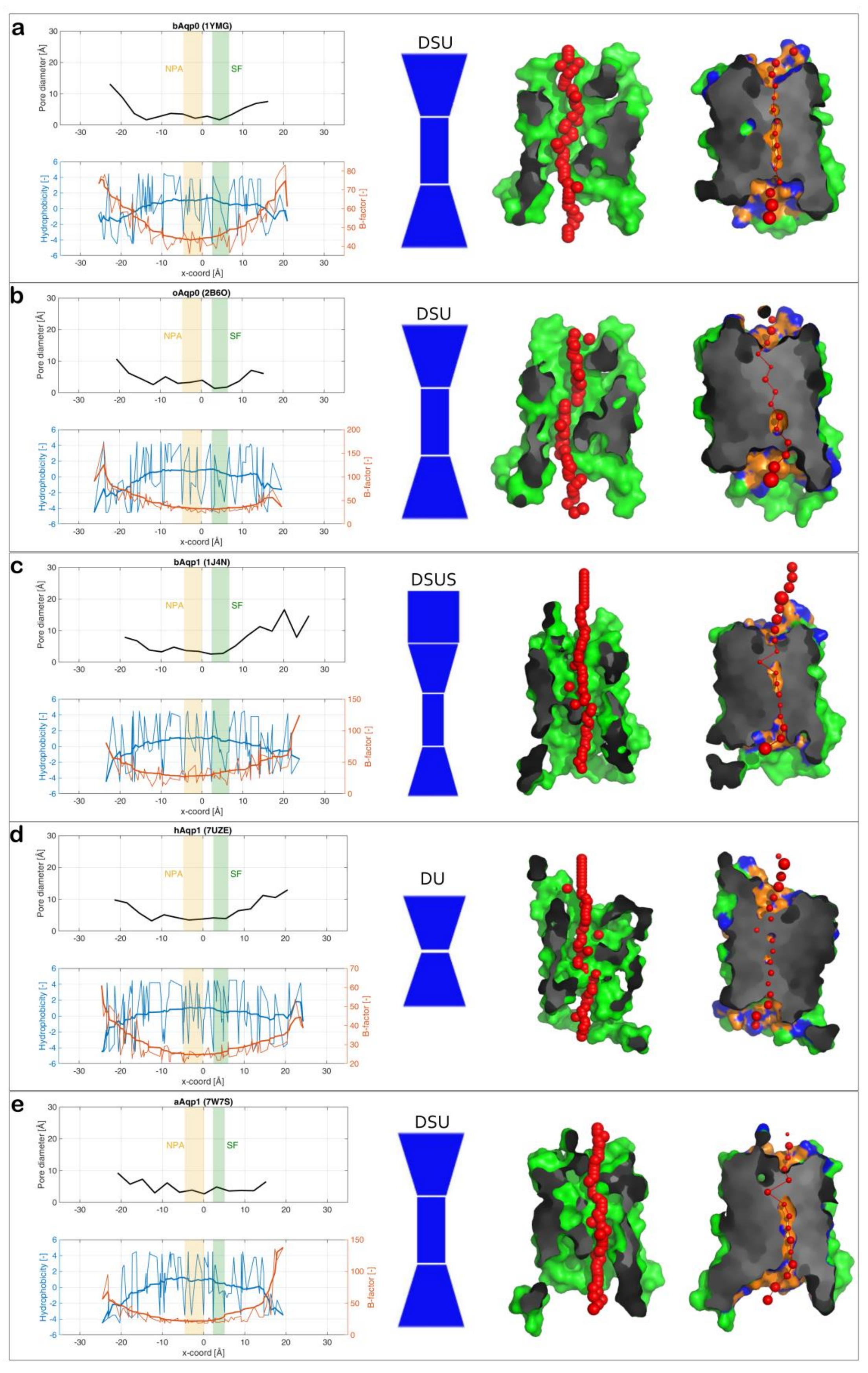

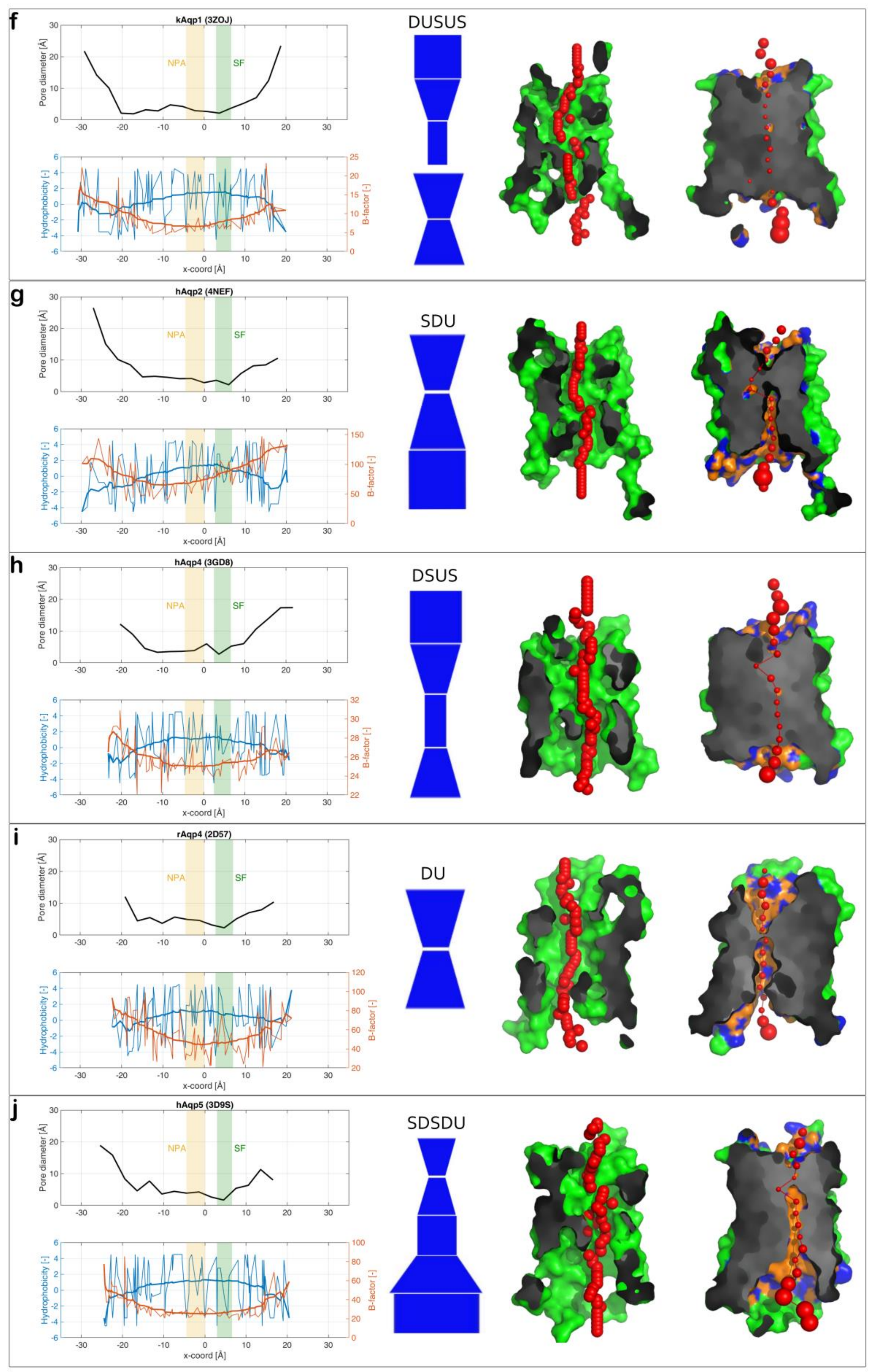

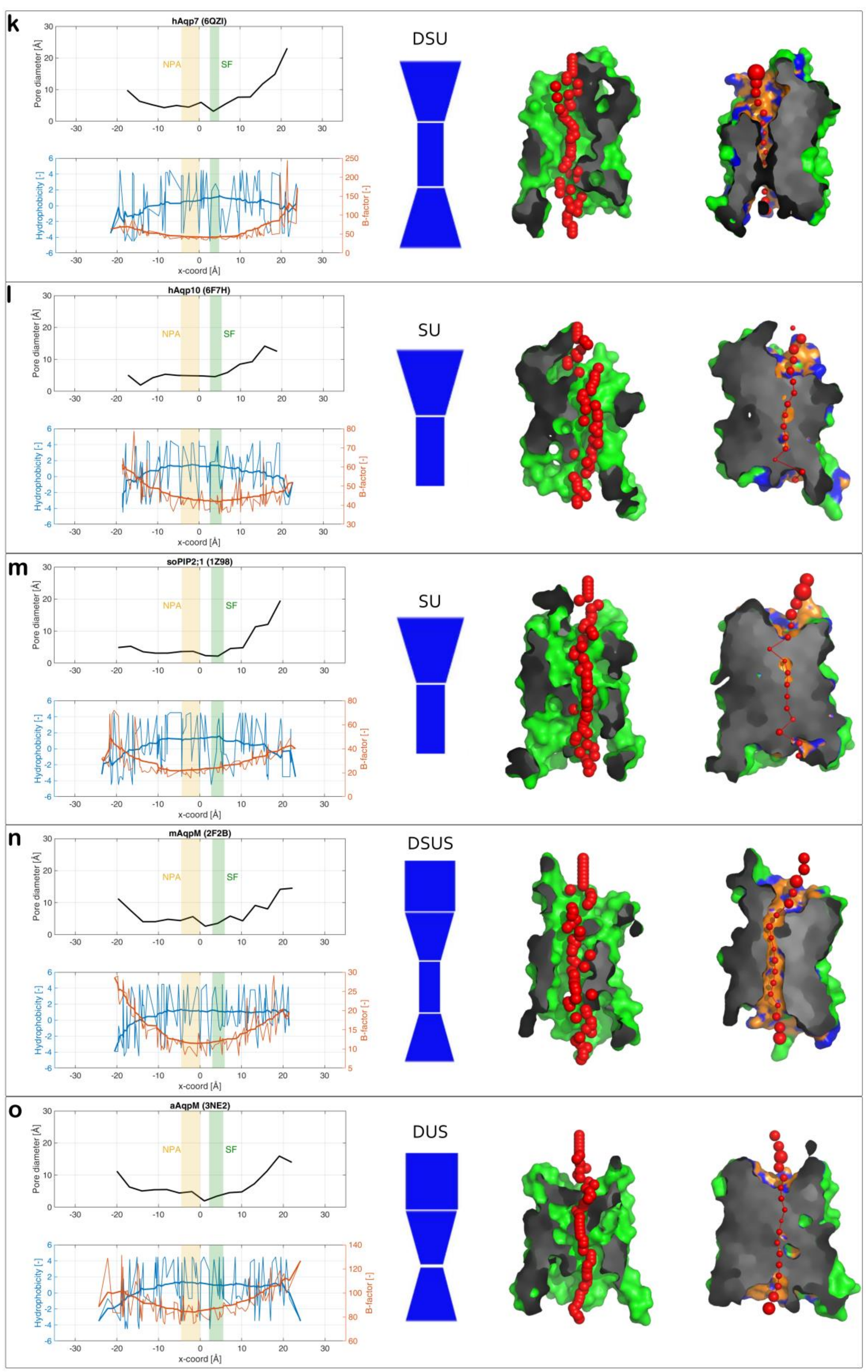

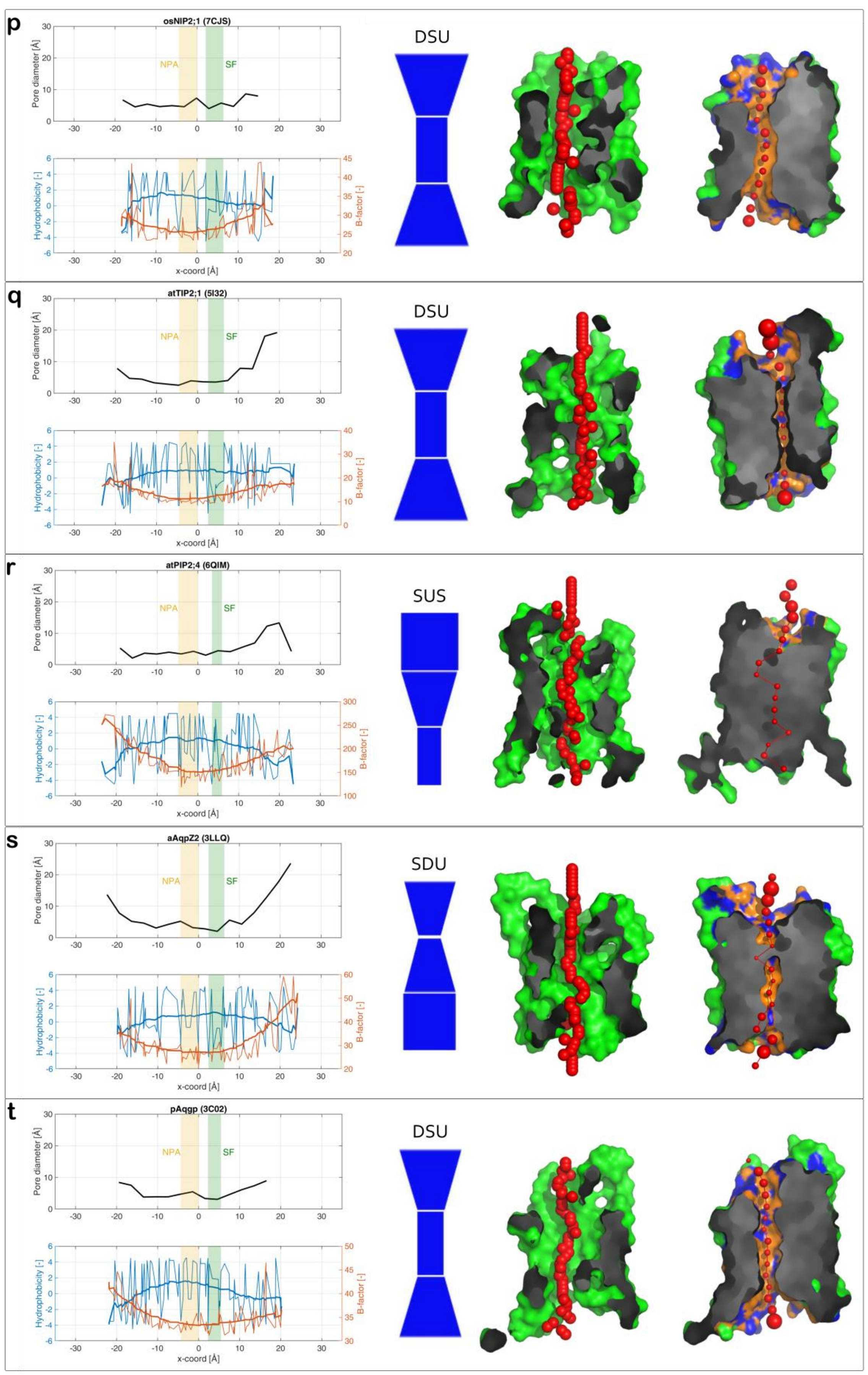

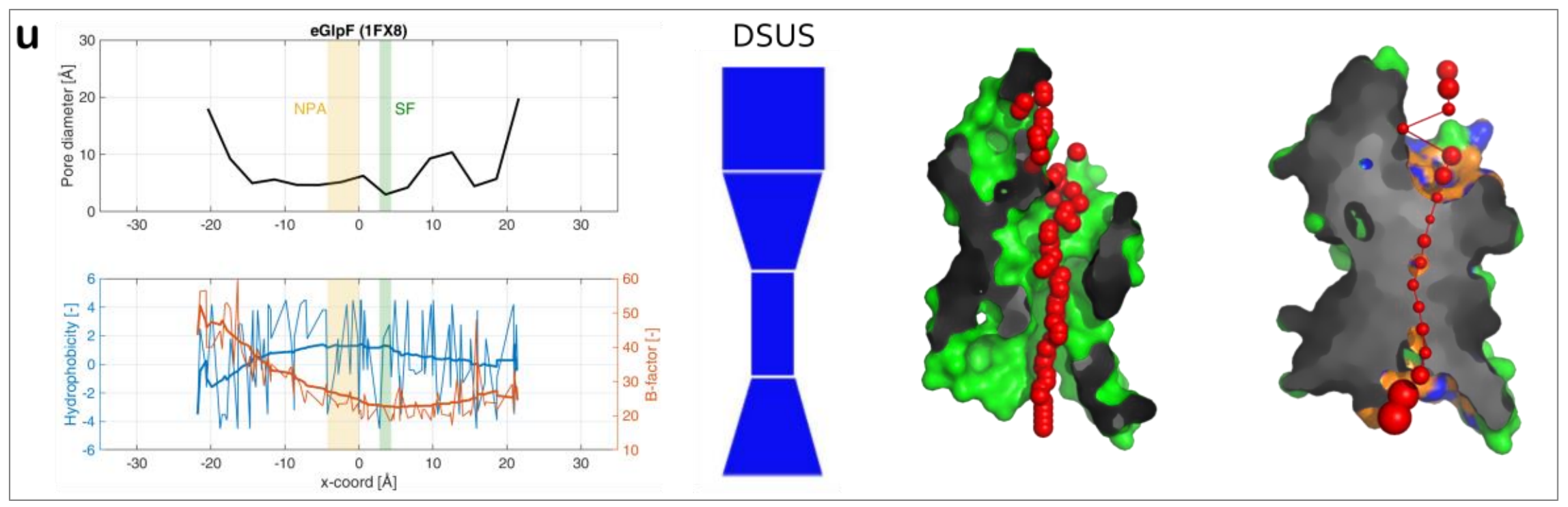

**Figure S1. Pore properties profiles and geometrical shape for all the AQ(G)Ps listed in Table 1 of the main text, except for *E. coli* AqpZ included in the main article.** Left top panels: diameter profile along the pore principal axis (*z*-coordinate) at 3Å step as calculated by the PoreWalker tool [56]. Left bottom panels: per-residue hydrophobicity index (blue) and B-factor (orange) of amino acids in the pore lining along the pore *z*-coordinate, as calculated by the Discovery Studio (BIOVIA, Dassault Systèmes) [96] and as obtained from the crystal PDB files, respectively. The thicker lines are smoothed curves (noise removed). Vertical bands depicted show the *z*-coordinates ranges where the NPA (orange) and SF (green) motifs are located. The blue funnel-like form at the center is a 2D projection of the predicted pore shape as calculated by variations of diameter values along the pore: the bottom of the pore axis (i.e., the lowest *z*-coordinate) correspond to the bottom of the stack. In the letter code on the top of the form, 'D' indicates a conical frustum generated by decreasing pore diameter values, i.e. the diameter of the lower base of the frustum is bigger than the diameter of its upper circle, 'U' indicates the opposite conical frustum, i.e., lower base diameter smaller than upper base diameter, and 'S' indicates a cylinder (the code ordering from bottom to top). The two longitudinal cross-sectional images are selected cuts along the pore's *z*-coordinate issued by PoreWalker, where the pore regularity is shown in the first cross-sectional image through red spheres located at the centre of the pore every 1Å step, whereas the pore dimension is represented in the second cross-sectional image by red spheres—at the center of the pore at 3Å step—, whose sizes are proportional to the pore diameter calculated at that point. In the latter, orange and blue areas show pore-lining atoms and corresponding residues, respectively. (a) bAQP0 (PDB ID 1YMG); (b) oAQP0 (PDB ID 2B6O); (c) bAQP1 (PDB ID 14JN); (d) hAQP1 (PDB ID 7UZE); (e) aAqp1 (PDB ID 7W7S); (f) kAqp1 (PDB ID 3ZOJ); (g) hAQP2 (PDB ID 4NEF); (h) hAQP4 (PDB ID 3GD8); (i) rAQP4 (PDB ID 2D57); (j) hAQP5 (PDB ID 3D9S); (k) hAQP7 (PDB ID 6QZI); (l) hAQP10 (PDB ID 6F7H); (m) SoPIP2;1 (PDB ID 1Z98); (n) mAqpM (PDB ID 2F2B); (o) aAqpM (PDB ID 3NE2); (p) OsNIP2;1 (PDB ID 7CJS); (q) AtTIP2;1 (PDB ID 5I32); (r) AtPIP2;4 (PDB ID 6QIM); (s) aAqpZ2 (PDB ID 3LLQ); (t) pAgqp (PDB ID 3C02); (u) eGlpF (PDB ID 1FX8). **Notes:** Aquaporin full names and their source organisms are given in Table 1 of the main text. Amino acids in the pore lining are reported in the PoreWalker's output files (**Supplementary Files**). The *z*-coordinates in the panels at the left (diameter, hydrophobicity index and B-factors profiles) were shifted from the original ones issued by PoreWalker, to place the beta carbon (βC) of the asparagine (N) in the second NPA motif at the center of the *z*-coordinate system (*z*=0). This ensures that all the AQ(G)Ps are aligned to have the same reference *z*-coordinate.

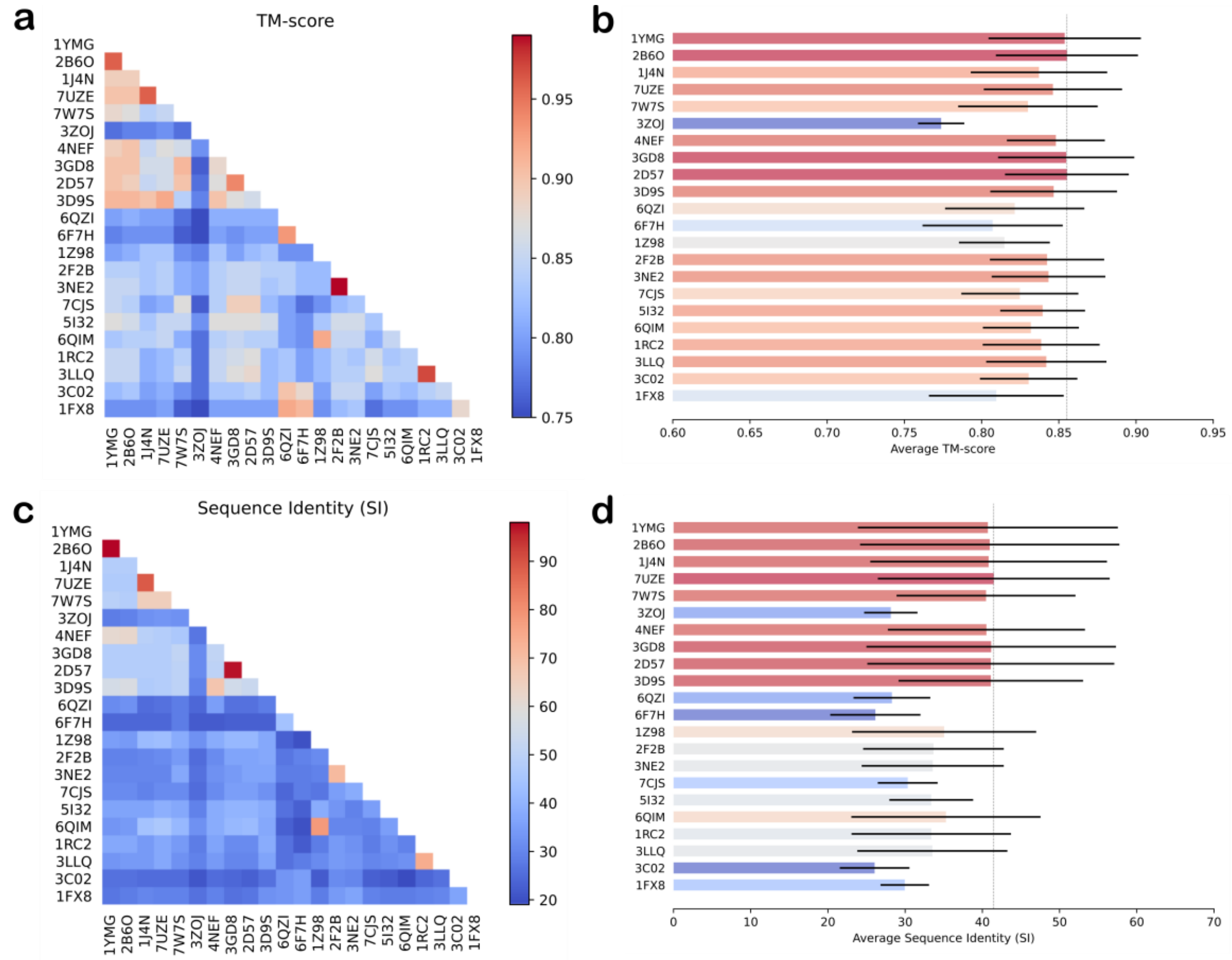

**Figure S2. Structural and sequence comparisons of AQ(G)Ps.** (a) TM-score-based pairwise map. (b) Per structure averaged TM-score against all other structures (bars). (c) Sequence identity (SI) pairwise map. (d) Per structure averaged SI against all other structures (bars). A min-max normalization is adopted for the bars coloring. The black lines at the right-hand side of the bars represent the standard deviations, whereas the dashed vertical lines indicate the maximum averages among all the structures. The initial chain of the tetrameric AQ(G)Ps' PDB structures was taken for the calculations. The corresponding AQ(G)Ps of the compared structures (PDBs) are given in Table 1 of the main text.

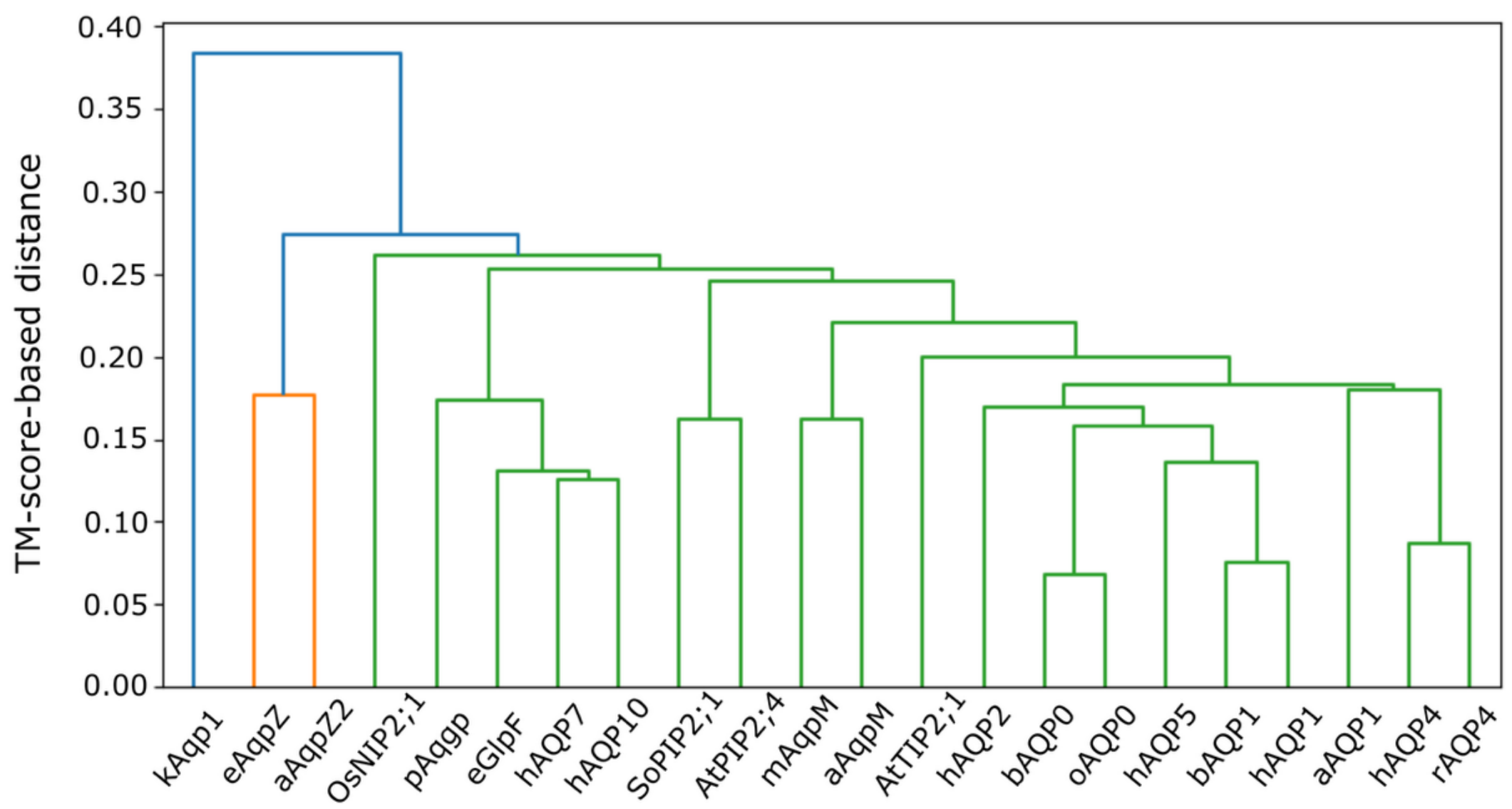

**Figure S3. Hierarchical clustering based on the pairwise TM-score of AQ(G)Ps.** The PDB IDs of the structures used are those reported in Table 1 of the main text. The single linkage method from the Python SciPy library (v.1.11.4) [97] was used for the clustering.

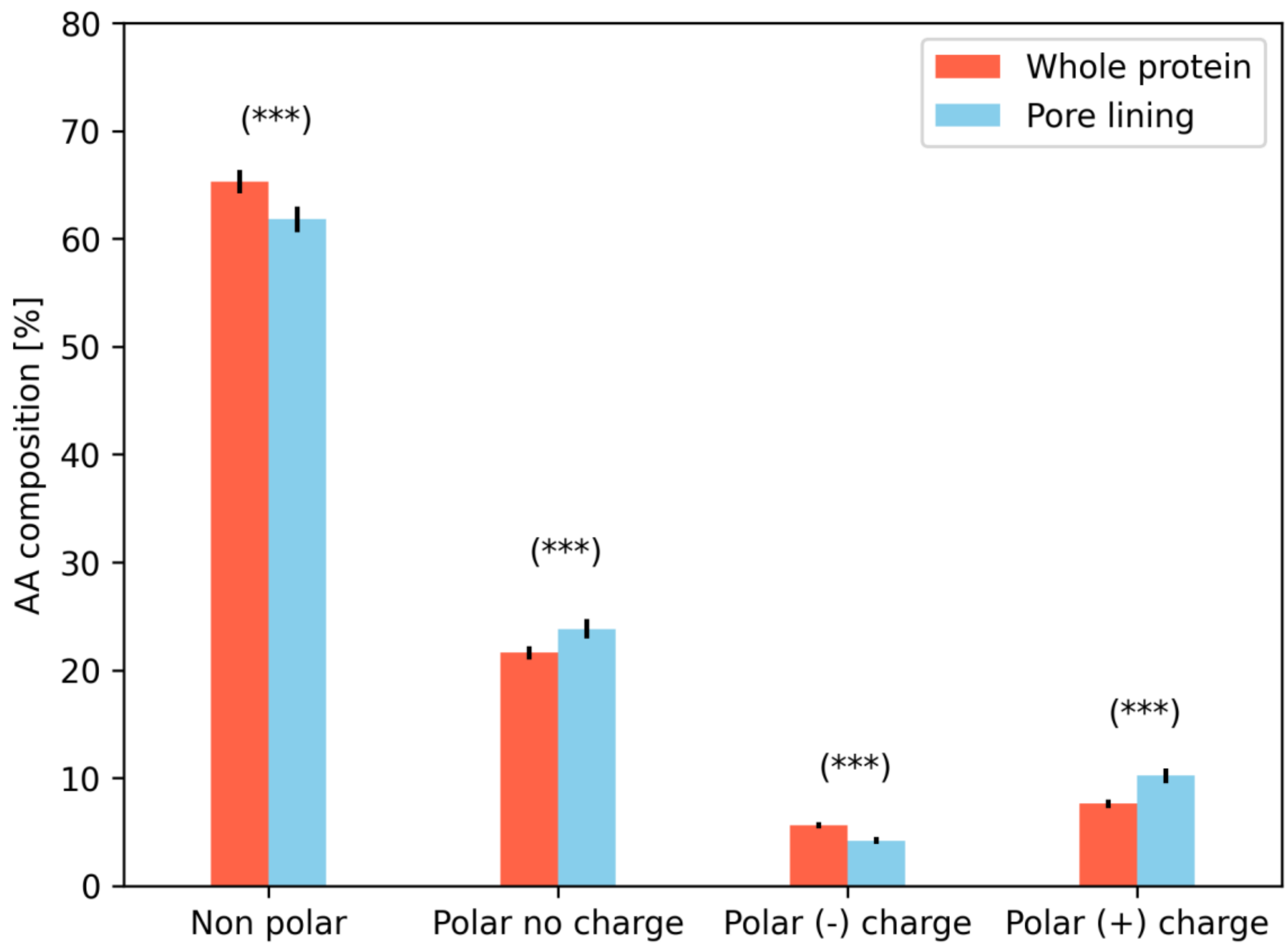

**Figure S4. Comparative depiction of the per-amino-acid type composition (in %) both in the whole protein and the pore lining across the AQ(G)Ps investigated.** The statistical significance (p-value < 0.01) of a t-test for mean differences is shown on each pair of compared bars. Data on the whole protein composition were obtained from the canonical aquaporin sequences retrieved from UniProt database [98]. Statistics on the pore lining composition were calculated based on the information given by PoreWalker [56]. FASTA files with the protein sequences used are provided as **Supplementary Files** along with the PoreWalker output files, which contain the pore shape, size and residue-level lining composition.

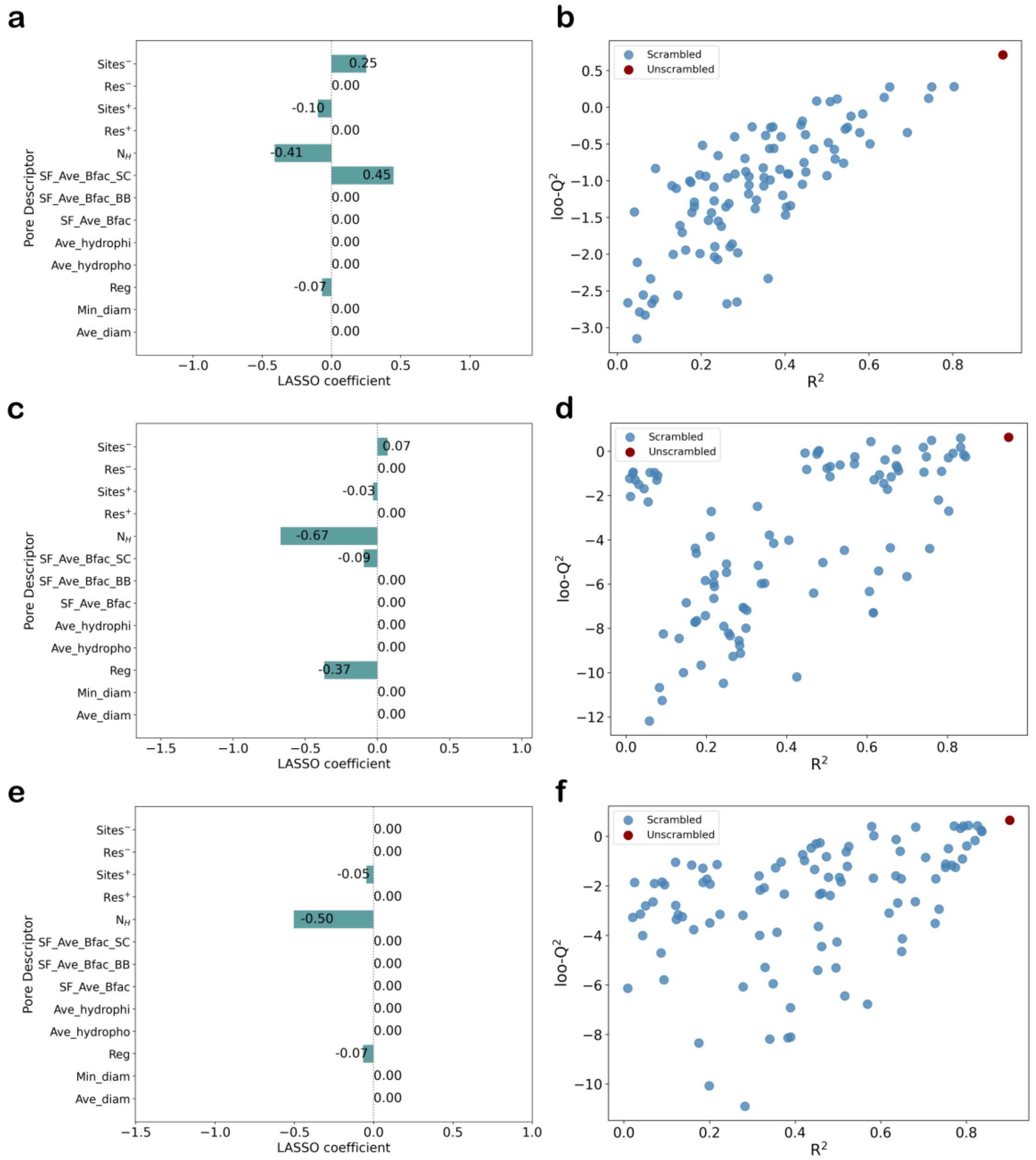

**Figure S5. Final LASSO coefficients and *y*-scramblings analyses for the best-performing QSAR models.** Panels a, c and e show bar plots of the final LASSO coefficients for models 1.3, 2.3a and 2.3b (see Table 3 of the main text), respectively, after applying the feature selection procedure. Coefficients were estimated using α = 0.1, 0.5, and 1.0 for models 1.3, 2.3a, and 2.3b, respectively, with 1,000,000 iterations in all cases. Panels b, d and f present the corresponding y-scrambling plots, showing leave-one-out predictive performance (loo-$Q^2$) as a function of the coefficient of determination ($R^2$) for 100 scrambled models (blue dots) compared with the original (unscrambled) model (red dot).

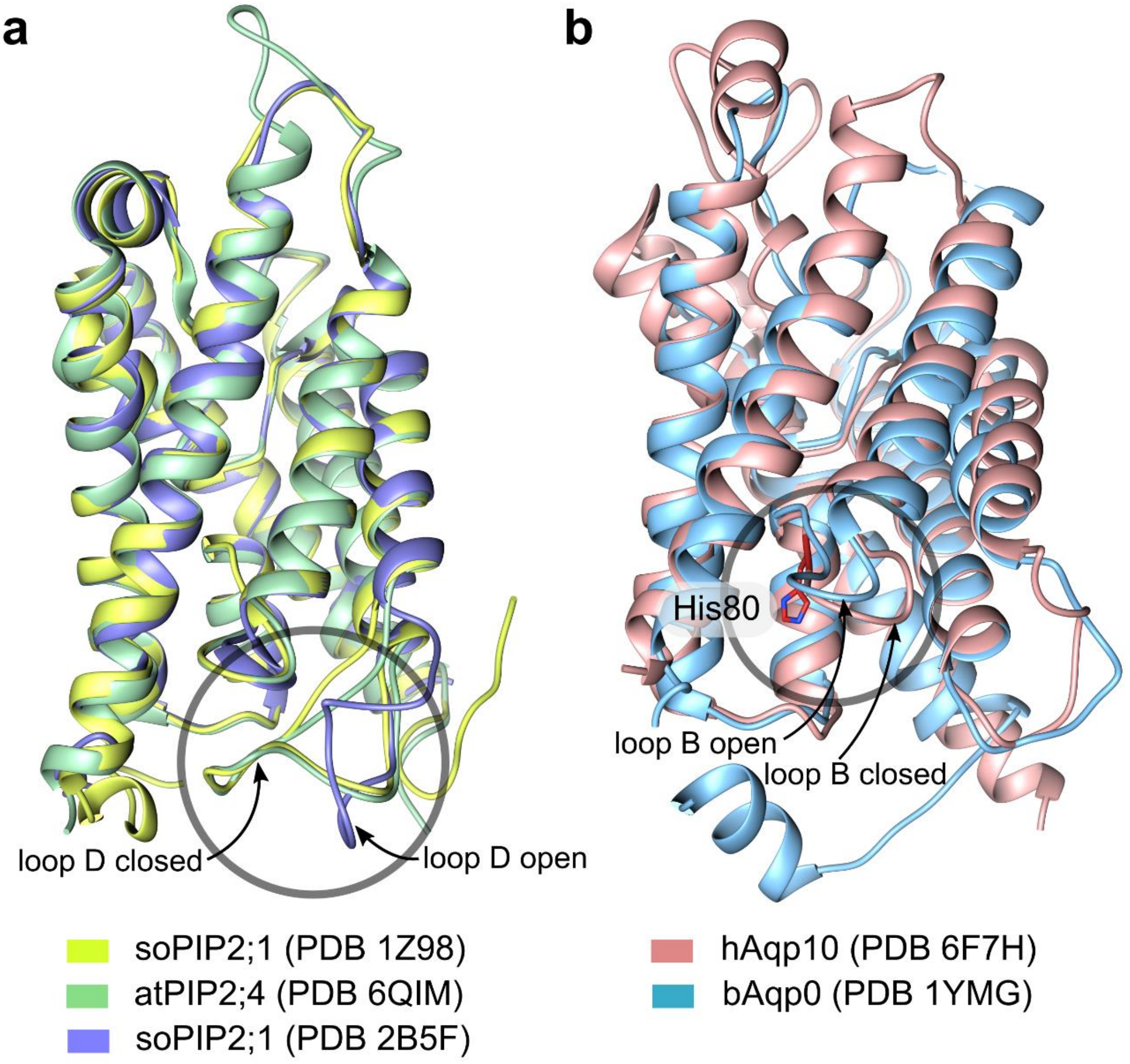

**Figure S6. Structure comparison of AQPs on pore gating molecular events.** (a) Overlapping of *Spinacia oleracea* PIP2;1, in its closed (PDB ID 1Z98 [45]) and open conformations (PDB ID 2B5F [45]), with the closed conformation of *Arabidopsis thaliana* PIP2;4 (PDB ID 6QIM [47]) aquaporin. Highlighted (circled region) are the main conformational differences in the cytoplasmic loop D that describe the 'pore-gating' event. (b) Overlapping hAQP10 (PDB ID 6F7H [48]) with bAQP0 (PDB ID 1YMG [34]) to highlight (circled region) the gated (closed) conformation of loop B in the former. His80, described as the key residue for the pH-dependent loop-B conformational change (pH-sensor [48]), is depicted (sticks).

## SUPPLEMENTARY REFERENCES